\pdfoutput=1
\documentclass[12pt,letterpaper,floatfix]{article}
\usepackage{amsmath,amssymb,amsthm,bbm,color}
\usepackage{graphicx}
\usepackage{epsfig}
\usepackage{euscript}
\usepackage{pslatex}
\usepackage{fancybox}
\usepackage{enumerate}
\usepackage{widetext}
\usepackage{subfigure}

\newcommand{\Meu}{\EuScript{M}}

\newcommand{\KK}{${\cal KK}$}

\newcommand{\sgn}{\mathop{\mathrm{sgn}}}
\def\st{\hbox{}} 
\usepackage[figuresright]{rotating}

\begin{document}
\begin{titlepage}


\begin{center}
\begin{flushleft}
{\small \bf BU-HEPP-19-10, Dec. 2019}
\end{flushleft}
\vspace{18mm}

{\bf\Large Initial-Final-Interference and Initial-State-Radiation Effects for Z/$\gamma^*$ Drell-Yan Observables using \KK{MC}-hh}\\
\vspace{2mm}

{S. Jadach$^a$, ~B.F.L. Ward$^b$,~Z. W\c{a}s$^a$,~S.A. Yost$^c$}\\
{$^a$Institute of Nuclear Physics Polish Academy of Sciences, Cracow, PL}\\
{$^b$Baylor University, Waco, TX, USA}\\
{$^c$The Citadel, Charleston, SC, USA}\\
\end{center}
\centerline{\bf Abstract}
Continuing with our investigations of the expected sizes of multiple photon radiative effects in heavy gauge boson production with decay to charged lepton pairs in the context of the precision physics of the LHC,  using \KK{MC}-hh 4.22 we consider IFI (initial-final interference) and ISR (initial-state radiation) effects for specific Z/$\gamma^*$ Drell-Yan observables measured by the ATLAS and CMS Collaborations. With this version of \KK{MC}-hh, we have coherent exclusive exponentiation (CEEX) electroweak (EW) exact ${\cal O}(\alpha^2 L)$ corrections in a hadronic MC and control over the corresponding EW initial-final interference (IFI) effects as well, where the big log $L$ is $\ln\frac{Q^2}{m^2}$ when $m$ is the respective mass of the radiating charged particle which undergoes the momentum transfer $Q$. Specifically, we illustrate the interplay between cuts of the type used in the measurement of $A_{FB}$ and $A_4$ at the LHC and the sizes of the expected responses of the attendant higher order corrections. We find that there are per cent to per mille level effects in the initial-state radiation, fractional per mille level effects in the IFI and per mille level effects in the over-all ${\cal O}(\alpha^2 L)$ corrections that any treatment of EW corrections at the per mille level should consider. Our results are applicable to current LHC experimental data analyses.

\end{titlepage}


\section{Introduction}
The large data samples at 7TeV and the even larger data samples at 8TeV and 13TeV have ushered in the era of precision QCD$\otimes$EW physics for the LHC experiments for processes
such as single heavy gauge boson production with decay to lepton pairs. As examples, the ATLAS Collaboration has recently used
8 TeV data samples to measure the Drell-Yan angular coefficients~\cite{atlas-angcf-16} $A_0,\; \ldots,\; A_7$ as a prelude to a precision measurement of $\sin^2\theta_W$. These measurements have been followed by their use of  
 their 7 TeV data samples to measure the mass of the
W boson with the result~Ref.~\cite{atlasmw-17}: 
$$M_W     =   80370 \pm 7 (\text{stat.}) \pm 11 (\text{exp. syst.}) \pm 14 (\text{mod. syst.}) \text{MeV}\\
=   80370 \pm 19 \text{MeV},$$
where the first uncertainty is statistical, the second is the experimental systematic uncertainty,  and the third is the physics-modeling systematic uncertainty. As one of the most precise single measurements of $M_W$~\cite{cdf-d0} the result bodes well, given the remaining data samples that have yet to be analyzed, for a new level of precision in the observable $M_W$ as well as other EW observables in LHC physics. High precision in measurements of lepton directions in detectors of hadron colliders is the essential feature, see eg. Ref.~\cite{d0-a}.
\par
From the error budget the ATLAS measurement of $M_W$ we see the importance of the modeling systematic error as it is the largest contribution with a value of $14$ MeV. Given that the corresponding statistical error will be reduced by a factor of $\sim 4$ when all of the available data are analyzed, it is necessary to reduce the large modelling error in kind as much as possible. We note that, in the measurement of $M_W$ by the ATLAS Collaboration~\cite{atlasmw-17} the W production and decay systematics are estimated by comparison with the analogous systematics for the $Z/\gamma*$ production and decays. The latter systematics are impacted by the uncertainty on the corresponding EW corrections.\par
We have discussed the sizes of the various relevant EW corrections on the observables such as the Z/$\gamma^*\; p_T$, the lepton $p_T$ and di-lepton invariant mass in Ref.~\cite{kkmchh1}. What we have found can be illustrated by a comparison of the results, with ATLAS cuts, from {\KK}MC-hh for the lepton $p_T$ spectrum in Fig.~1 in ~\cite{kkmchh1} with the ATLAS ratio plot between their data and the best theory predictions which they employ as presented  in their Fig.~15 in Ref.~\cite{atlasmw-17}.
If we look at the effect of the ISR on this spectrum as predicted by 
{\KK}MC-hh we see that it agrees with the fact that the ATLAS data are about 1-2\% above the theory which they use at the low $p_T$ end of the respective plot and a similar amount below the theory at the high end of that plot. This is expected because the theory used by ATLAS, which features the QED ISR from the respective Pythia~\cite{py8} shower, does not have 
the full effect of the ISR from the transverse degrees of freedom for the radiation featured in the respective {\KK}MC-hh predictions. \par
Similarly, in the ATLAS and CMS measurements of the angular coefficients~\cite{atlas-angcf-16,cms-ang-coeff-15}, the systematics of the respective modelling errors are impacted by the uncertainty on the respective EW corrections. In what follows, we will explore to what extent the various aspects of the EW corrections interplay with the ATLAS-type cuts and the method of application of the corrections.\par
A key issue in this interplay is the role of phase space competition between photons and gluons in the multiple photon and multiple gluon processes under study here. Specifically, in Ref.~\cite{sjos-annecy:1991} in the context of FSR (final-state radiation) at LEP, it was shown that the competition between parton shower gluons and parton shower photons led to  considerable reduction in the available phase space for photons when the QED and QCD showers were interleaved relative to the situation in which the two showers are not interleaved. A natural question to ask is whether we have to take such a reduction into account in our calculations with {\KK}MC-hh for the respective Z/$\gamma^*$ observables which we study? The key point in the results in Ref.~\cite{sjos-annecy:1991} is that in the infrared regime, with energy fraction $\lesssim 0.1$, there is essentially no effect of phase space competition between the interleaved gluon and photon showers. This is important because in {\KK}MC-hh we resum the infrared regime to all orders in $\alpha$ in the presence of hard photon residuals. The hard photon residuals are separated in  space-time from the gluon shower quanta in the standard inside-out cascade, so that there is also no phase space competition between the hard photon residuals and the shower gluons. What we can have is a phase space competition between the hard photon residuals and the hard gluons in our processes, where the first such effects occur at ${\cal O}(\alpha\alpha_s)$. The size of such hard non-factorizable two-loop effects has been studied in Ref.~\cite{dittmr-1,dittmr-2} and it is expected to be below the level of precision in the studies we present here. 
\par
We should also call attention to the studies  in Refs.~\cite{wack1,wack2,ditt1,fulv1,ditt2} and  in Ref.~\cite{vicini-wack:2016} on the expected sizes of the EW corrections in LHC observables. We the detailed relationship between our {\KK}MC-hh results and those in these latter references will be addressed elsewhere~\cite{elswh}\footnote{This relationship is part of the ongoing studies in the CERN LPCC EW Precision Subgroup Meetings: https://lpcc.web.cern.ch/electroweak-precision-measurements-lhc-wg~\cite{froid:2019}.}. 
For neutral current Drell-Yan processes, Herwig~\cite{HERWIG}, Pythia~\cite{py8,Pythia}, Herwig++~\cite{hwg++} and Sherpa~\cite{sherpa} have featured QED radiative effects in the context of parton showers: the leading-log QED shower is implemented in Herwg, Herwig++, Pythia and Sherpa and final-state YFS~\cite{yfs:1961} exponentiated radiation for decays is implemented in Herwig++ and in Sherpa, where in Sherpa the NLO EW and QED NNLO exact YFS residuals (see Ref.~\cite{yfs:1961} for their definition)  have been realized~\cite{sherpa3} in this context. In Refs.~\cite{sherpa2,recola,opnlps}, Sherpa, Recola and OpenLoops authors have also made available exact ${\cal O}(\alpha)$  EW corrections and exact NLO QCD corrections to such Drell-Yan processes as an option with parton showers. In the Powheg framework, the corresponding exact ${\cal O}(\alpha)$  EW corrections and exact NLO QCD corrections are realized as presented in Ref.~\cite{powheg1}. We additionally observe that  SANC~\cite{sanc} features NLO EW and NLO QCD corrections to neutral current Drell-Yan processes and that FEWZ~\cite{fewz} realizes the exact ${\cal O}(\alpha)$  EW corrections along with its  exact NNLO QCD corrections to such processes. We would again note the exact ${\cal O}(\alpha\alpha_s)$ non-factorizable corrections to the neutral current Drell-Yan process already referenced in Refs.~\cite{dittmr-1,dittmr-2}, which are available, along with the NLO QCD and NLO EW corrections, in the MC integrator program RADY. In the connection of mixed ${\cal O}(\alpha\alpha_s)$ corrections, we finally note the studies in Refs. ~\cite{delto1,bonciani1} for on-shell Z production.\par

The paper is organized as follows. In the next section we give a brief review of the physics  in the {\KK}MC-hh MC, as it is still not a generally familiar.
In Section 3 we illustrate the effect of the EW corrections in {\KK}MC-hh in the context of the acceptance used by ATLAS in their studies of the angular coefficient $A_4$ and $A_{FB}$
in single $Z/\gamma^*$ events with decays to lepton pairs in Ref.~\cite{atlas-angcf-16}. In this section, we make contact with the studies in Ref.~\cite{vicini-wack:2016}. 
In Section 4, we summarize our findings.\par

\section{The Physics in {\KK}MC-hh}
In {\KK}MC-hh we combine the exact amplitude-based  CEEX/EEX YFS MC approach to EW higher order corrections pioneered in Refs.~\cite{Jadach:1993yv,Jadach:1999vf,Jadach:2000ir,Jadach:2013aha} and the QCD parton shower hadron MC approach pioneered in Refs.~\cite{sjos-sh,HERWIG}. Here, EEX denotes exclusive exponentiation as originally formulated by Yennie, Frautschi and Suura (YFS) in Ref.~\cite{yfs:1961}. In the discussion which follows, we will use the Herwig6.5~\cite{HERWIG} MC for the parton shower realization although the use of any parton shower MC which accepts LHE~\cite{lhe-format} input is allowed in {\KK}MC-hh studies. We now give a brief review of the physics in {\KK}MC-hh.\par

Since it is still not generally used, we recall the master formula for the CEEX realization of the higher corrections to the SM~\cite{SM1,SM2,SM3,SM4} EW theory. 
For clarity, we note that the CEEX realization is amplitude level coherent exclusive exponentiation whereas the EEX realization is exclusive exponentiation at the squared amplitude level. For the purpose of illustration, let us consider the prototypical process 
$q\bar{q}\rightarrow \ell\bar{\ell}+n\gamma, \; q=u,d,s,c,b,t,\ell=e,\mu,\tau,\nu_e,\nu_\mu,\nu_\tau.$  For this process, we have the cross section formula
\begin{equation}
\sigma =\frac{1}{\text{flux}}\sum_{n=0}^{\infty}\int d\text{LIPS}_{n+2}\; \rho_A^{(n)}(\{p\},\{k\}),
\label{eqn-hw2.1-1}
\end{equation}
where $\text{LIPS}_{n+2}$ denotes Lorentz-invariant phase-space for $n+2$ particles, $A=\text{CEEX},\;\text{EEX}$, the incoming and outgoing fermion momenta are abbreviated as $\{p\}$ and the $n$ photon momenta are denoted by $\{k\}$.
Thanks to use of conformal symmetry, full $2+n$ body  phase space is covered without any
approximations. The respective algorithm's details are covered in Ref.~\cite{Jadach:1999vf}.
Specifically, we have from Refs.~\cite{Jadach:2000ir,Jadach:1999vf,kkmchh} that 
\begin{equation}
\rho_{\text{CEEX}}^{(n)}(\{p\},\{k\})=\frac{1}{n!}e^{Y(\Omega;\{p\})}\bar{\Theta}(\Omega)\frac{1}{4}\sum_{\text{helicities}\;{\{\lambda\},\{\mu\}}}
\left|\Meu\left(\st^{\{p\}}_{\{\lambda\}}\st^{\{k\}}_{\{\mu\}}\right)\right|^2.
\label{eqn-hw2.1-2}
\end{equation}
(For the corresponding formula for the $A=\text{EEX}$ case we refer the reader to  Refs.~\cite{Jadach:2000ir,Jadach:1999vf}.) Here,  $Y(\Omega;\{p\})$ is the YFS infrared exponent. The respective infrared integration limits are specified by the region $\Omega$ and its characteristic function
$\Theta(\Omega,k)$ for a photon of energy $k$, with $\bar\Theta(\Omega;k)=1-\Theta(\Omega,k)$ and $$\bar\Theta(\Omega)=\prod_{i=1}^{n}\bar\Theta(\Omega,k_i).$$
The definitions of the latter functions as well as the CEEX amplitudes $\{\Meu\}$ are given in Refs.~\cite{Jadach:1999vf,Jadach:2000ir,Jadach:2013aha}. 
{\KK}MC-hh obtains from \KK MC 4.22 the exact
${\cal O}(\alpha)$ EW corrections implemented using the  DIZET 6.21 EW library from the semi-analytical
program ZFITTER~\cite{zfitter1,zfitter6:1999}. The respective implementation is described in Ref.~\cite{Jadach:2000ir} so that we do not repeat it here.
In {\KK}MC-hh, the CEEX amplitudes $\{\Meu\}$ in (\ref{eqn-hw2.1-2}) are exact in ${\cal O}(\alpha^2 L^2, \alpha^2L)$ in the sense that all terms in the respective cross section at orders ${\cal O}(\alpha^0),\;{\cal O}(\alpha),\; {\cal O}(\alpha L),\;{\cal O}(\alpha^2 L), \; \text{and}\; {\cal O}(\alpha^2 L^2)$ are all included in our result for that cross section.
Here the big log is $L=\ln\frac{Q^2}{m^2}$ where
$Q$ is the respective hard 4-momentum transfer. In our case, the light quark masses and the charged lepton masses will determine $m$, depending on the specific process
under consideration. We follow Ref.~\cite{mstw-mass} and use the current quark masses~\cite{PDG:2016} $m_u = 2.2 \text{MeV}.\; m_d = 4.7 \text{MeV}, \;  m_s = 0.150\text{GeV}, \; m_c = 1.2 \text{GeV},\;  m_b = 4.6\text{GeV}, \text{and} m_t = 173.5 \text{GeV}$\footnote{See Ref.~\cite{kkmchh1} for a discussion of the uncertainty of our results due to realistic uncertainties on our values of the current quark masses.}.
\par

The realization of the parton shower MC approach proceeds via the standard Drell-Yan formula for the process $pp\rightarrow Z/\gamma^*+X\rightarrow \ell\bar{\ell}+X'$, $\ell = e^-,\mu^-$:
\begin{equation}
\sigma_{\text{DY}}=\int dx_1dx_2\sum_i f_i(x_1)f_{\bar{i}}(x_2)\sigma_{\text{DY},i\bar{i}}(Q^2)\delta(Q^2-x_1x_2s),
\label{eqn-hw2.1-3}
\end{equation}
where the subprocess cross section for the $i$-th $q\bar{q}$ annihilation with $\hat{s}=Q^2$ when the pp cms energy squared is $s$
is denoted in a conventional notation for parton densities $\{f_j\}$. Here, $\sigma_{\text{DY},i\bar{i}}(Q^2)$ is given by the right-hand-side of Eq.~\ref{eqn-hw2.1-1}, realized by Monte Carlo methods as explained in Refs.~\cite{Jadach:2013aha,kkmchh}. {\KK}MC-hh receives multiple gluon radiation, for a given QCD parton shower MC, via the backward evolution~\cite{sjos-sh} 
for the densities as specified in
(\ref{eqn-hw2.1-3}). This backward evolution then also affords {\KK}MC-hh the hadronization for the attendant shower.
While we use here the Herwig6.5 shower MC for this phase of the event generation, we note, again, that, as the Les Houches Accord format is also available for the hard processes generated in {\KK}MC-hh before the shower, all shower MC's which use that format can be used for the shower/hadronization part of the simulation.\par

\section{CEEX Exact ${\cal O}(\alpha^2L)$ EW IFI and ISR Effects from {\KK}MC-hh for the ATLAS Acceptance for Z/$\gamma^*$ Drell-Yan Observables}
As we have noted, in both the ATLAS and the CMS Collaborations measurements of the Z/$\gamma^*$ decay angular coefficients~\cite{atlas-angcf-16,cms-ang-coeff-15}, the systematics of the respective modelling errors are impacted by the uncertainty on the respective EW corrections. In this section we use the $Z/\gamma^*$ cuts that are typical of the systematics studies done by ATLAS in their angular coefficients and $m_W$ analysis, as motivated by our participation in the CERN Precision Studies Subgroup of the LPCC EWWG\footnote{https://lpcc.web.cern.ch/electroweak-precision-measurements-lhc-wg~\cite{froid:2019} -  note that several of the authors of the purely collinear QED PDF approaches presented in Ref.~\cite{vicini-wack:2016} are involved in these studies.}, to illustrate the size of the new 
higher order EW effects
in {\KK}MC-hh in the context of those cuts for the angular coefficients and related observables.\par

The ATLAS-type cuts on the $Z/\gamma^*$ production and decay to lepton pairs, as employed in Ref.~\cite{atlas-angcf-16,atlasmw-17}, which we use are as follows:
$$ 60\;\text{ GeV}<M_{\ell\ell}<116\; \text{GeV},\; P^{\ell\ell}_{T}< 30\;\text{GeV},\; $$
where both members of the decay lepton pair satisfy $$ P^{\ell}_{T}> 25\;\text{ GeV},\; |\eta_\ell| < 2.5. $$ 
Here, we have defined  $M_{\ell\ell}$ as the lepton pair invariant mass, $P^{\ell\ell}_{T}$ as the transverse momentum of the lepton pair,  $P^{\ell}_{T}$ as the transverse momentum of the lepton or anti-lepton $\ell$, and $\eta_\ell$ as the pseudorapidity of the lepton or anti-lepton $\ell$. \par

\begin{figure}[h]
\begin{center}
\setlength{\unitlength}{1in}
\begin{picture}(6,2.4)(0,0)
\put(1.0,0.2){\includegraphics[width=4in]{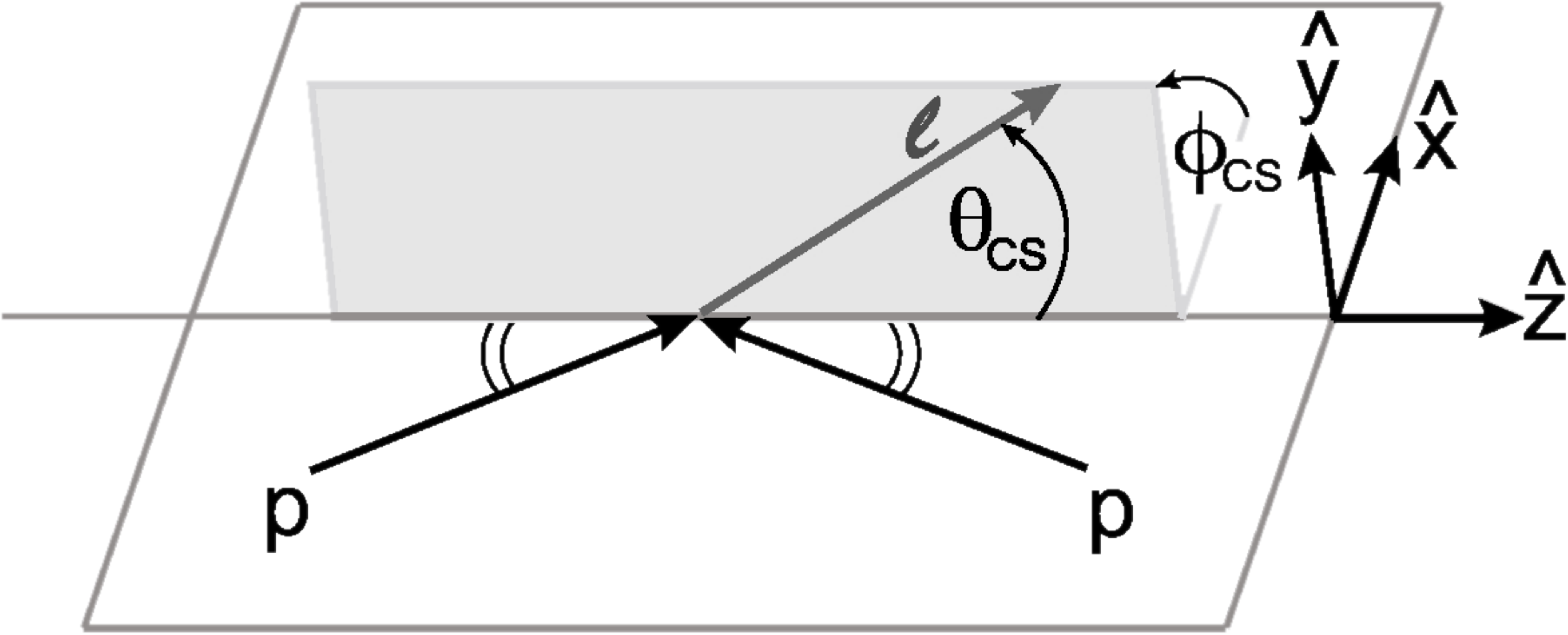}}
\end{picture}
\end{center}
\vspace{-10mm}
\caption{\baselineskip=11pt Kinematics of the lepton decay angles in the Collins-Soper frame~\cite{atlas-angcf-16,cs-frame}. The Collins-Soper angle is given by $cos\theta_{CS}=\sgn(P^z_{\ell\ell})\frac{p^+_{\ell1}p^-_{\ell2}-p^-_{\ell1}p^+_{\ell2}}{M_{\ell\ell}\sqrt{M_{\ell\ell}^2+P^2_{T\ell\ell}}}$ with $p^{\pm}=p^0\pm p^z$.$P_{\ell\ell}=p_{\ell1}+p_{\ell2}$, where $\ell_1=\ell^-\equiv \ell$ in the illustration and $\ell_2=\ell^+\equiv \bar{\ell}$. The laboratory z-axis may taken as that given in Ref.~\cite{atlas-1109.5141}.}
\label{fig-csfrme-kin-1}
\end{figure}

We start with the basic kinematics for the observables which we study, as it is shown in Fg.~\ref{fig-csfrme-kin-1}. 
We work in the Collins-Soper(CS) frame for the outgoing lepton pair with the Collins-Soper~\cite{cs-frame} polar and azimuthal angles $\theta_{CS},\; \phi_{CS}$ as shown in Fig.~\ref{fig-csfrme-kin-1}. Since we are interested in the systematics associated with the extraction of $\sin^2\theta_W$ from the respective data,
we will focus on the angular distribution for $\theta_{CS}$ and the observables $A_4$ and $A_{F B}$, which we define as 
$$A_4=\frac{4}{\sigma}\int\cos\theta_{CS}d\sigma=4<\cos\theta_{CS}>$$,
$$A_{FB}=\frac{1}{\sigma}\int\sgn(\cos\theta_{CS})d\sigma=<\sgn(\cos\theta_{CS})>$$
where we follow the notation of Ref.~\cite{mirkes-1} for the angular coefficients $A_i,\; i=0,\ldots, 7$ in the respective differential cross section $d\sigma(\theta_{CS},\phi_{CS})$ for the 
lepton in Fig.~\ref{fig-csfrme-kin-1}\footnote{Consideration of more radiative effects than those discussed explicitly in Ref.~\cite{mirkes-1} will lead, in general, to the introduction of coefficients beyond those in Ref.~\cite{mirkes-1} for higher $\ell$ spherical harmonics $Y^m_\ell$; as these are orthogonal to $Y^0_1$, our formula for $A_4$ is unaffected but the relationship between $A_4$ and $A_{FB}$ is affected as for example $Y^0_3$ has a forward-backward asymmetry.}.\par
As we have shown in Ref.~\cite{kkmchh1} QED ISR (initial-state radiation) enters the angular distributions at the level of several per mille and cannot be neglected. In what follows, we compare our exact ${\cal O}(\alpha^2 L)$ CEEX treatment of these effects in {\KK}MC-hh with their treatment in the QED-pdf approach as it is realized with the LuxQED~\cite{luxqed} formulation as realized in the NNPDF 3 1 NLO ($\alpha_s(M_Z)=0.118$) set~\cite{nnpdf3.1}. We expect that the two approaches should agree when effects are not sensitive to the photon transverse momentum $p_{\gamma,T}$.\par
Specifically, we turn first to comparisons featuring unshowered results using {\KK}MC-hh in which we have a sample of $5.7\times 10^9$ muon-pair events\footnote{In this paper we focus on muon pairs in the $Z/\gamma^*$ decays so that we avoid the issues of calorimetry for $e^+e^-$ pairs and of tau decays in $\tau^+\tau^-$ pairs. The latter two scenarios will be taken up  elsewhere~\cite{elswh}. Note that it is known that IFI is roughly the same for both bare and dressed FSR because
IFI comes mainly from photons in the middle of the angular range between IS and FS charges. Since the data in Ref.~\cite{atlas-angcf-16} are at the Born $Z$ level to facilitate combination of electron and muon data, our treatment of bare muons allows one to see the relative sizes of the ISR and IFI effects we calculate compared to the FSR used to arrive at Born Z results.} at 8 TeV. In the discussion of our results, "uncut/without cuts" means that no additional cuts beyond the muon-pair mass cut $ 60\;\text{ GeV}<M_{\ell\ell}<116$ are made whereas "cut/with cuts" means that the additional cuts $ P^{\ell}_{T}> 25\;\text{ GeV},\; |\eta_\ell| < 2.5$ are made on both members of the muon pair in conjunction with the cut $P^{\ell\ell}_{T}< 30$. Under these circumstances, we present four levels of photonic corrections:\\
1. Final-state radiation(FSR)  using {\KK}MC-hh with non-QED NNPDF3.1 NLO \\
2. FSR + initial-state radiation (ISR) using {\KK}MC-hh with non-QED NNPDF3.1 NLO \\
3. FSR + ISR + initial-final interference (IFI) radiative effects using {\KK}MC-hh with non-QED NNPDF3.1 NLO (the best {\KK}MC-hh result) \\
4. FSR + LuxQED using {\KK}MC-hh with QED NNPDF3.1 NLO.\\
The {\KK}MC-hh photonic corrections are calculated using CEEX exponentiation with exact
${\cal O}(\alpha^2L)$ residuals.\par
We present in Table~\ref{tab-1} results for the uncut and cut cross sections and for the $A_{FB}$ and $A_4$ observables for the four levels of photonic corrections as described. 
\begin{table}[h]
\caption{\text{Numerical Results}}
\centering
\scalebox{.85}{
\begin{tabular}{|c|c|c|c|c|c|c|}
\hline
&{\small No ISR}&{\small  LuxQED}&{\small {\KK}MC-hh ISR}&{\small \%(ISR-no ISR)}&{\small With IFI}&{\small \%(IFI - no IFI)}\\
\hline
{\small\text{Uncut} $\sigma (\text{pb})$ }&{\small 939.86(1)}&{\small 944.04(1)}&{\small 944.99(2)}&{\small 0.54597(2)}&{\small 944.91(2)}&{\small $-0.0089(4)$} \\
{\small\text{Cut} $\sigma (\text{pb})$}&{\small 439.10(1)}&{\small 440.93(1)}&{\small 442.36(1)}&{\small 0.74223(3)}&{\small 442.33(1)}&{\small$-0.0070(5)$}\\
\hline
&{\small No ISR}&{\small LuxQED}&{\small {\KK}MC-hh ISR}&{\small ISR- no ISR}&{\small With IFI}&{\small IFI - no IFI}\\
\hline
{\small$A_{FB}$}&{\small 0.01125(2)}&{\small 0.01145(2)}&{\small 0.01129(2)}&{\small$(3.9\pm2.8)\times10^{-5}$}&{\small 0.01132(2)}&{\small$(2.9\pm1.1)\times10^{-5}$}\\
{\small$A_4$}&{\small 0.06102(3)}&{\small 0.06131(3)}&{\small 0.06057(3)}&{\small$-(4.4\pm0.5)\times10^{-5}$}&{\small0.06102(3)}&{\small$(4.5\pm0.3)\times10^{-5}$}\\
\hline
\end{tabular}}
\label{tab-1}
\end{table}
For the uncut and cut cross sections, we see that {\KK}MC-hh
shows an ISR effect of a fraction greater than half of a percent. LuxQED shows a slightly smaller
effect, about 0.4\% for each cross section. {\KK}MC-hh shows an IFI effect below 0.1\%. On the the angular coefficients the ISR and IFI effects are both on the order of
$10^{-5}$ in {\KK}MC-hh. 
LuxQED gives a somewhat bigger ISR effect in this case, on the order of $10^{-4}$.\par
\begin{figure}[h]
\begin{center}
\setlength{\unitlength}{1in}
\begin{picture}(6.5,4.7)(0,0)
\put(0.2,2.5){\includegraphics[width=3.2in,height=2.0in]{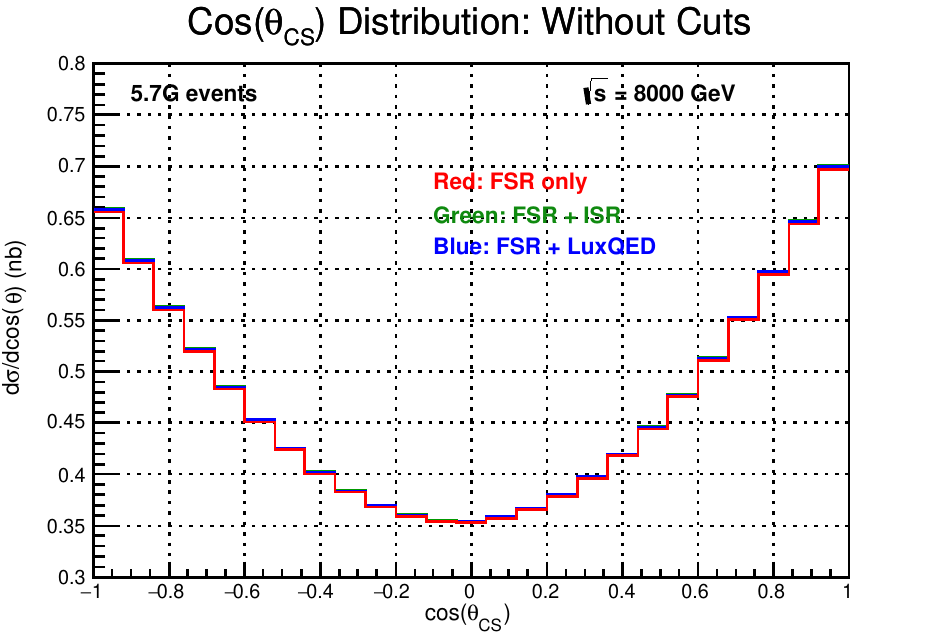}}
\put(3.2,2.5){\includegraphics[width=3.2in,height=2.0in]{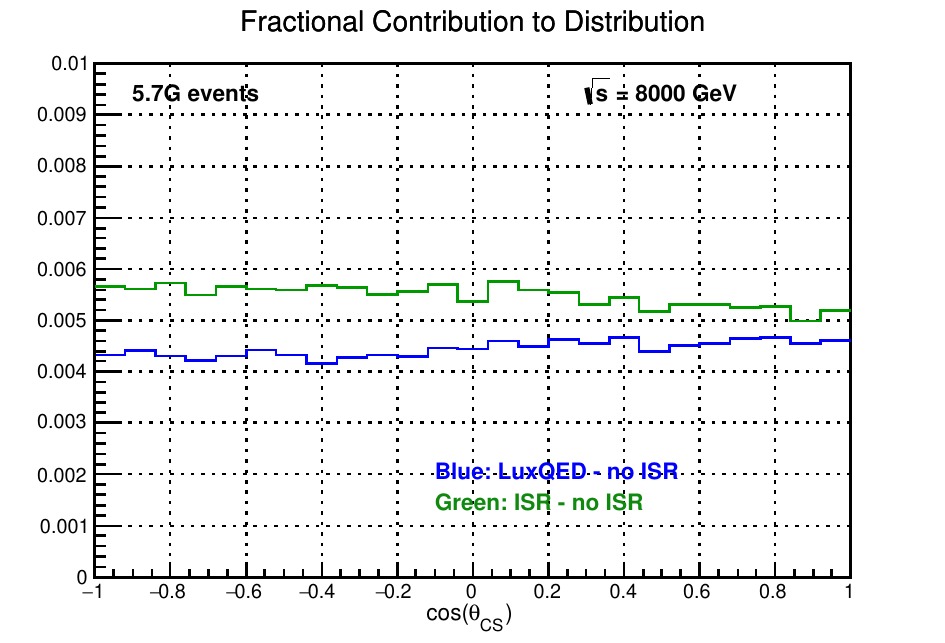}}
\put(0.2,.2){\includegraphics[width=3.2in,height=2.0in]{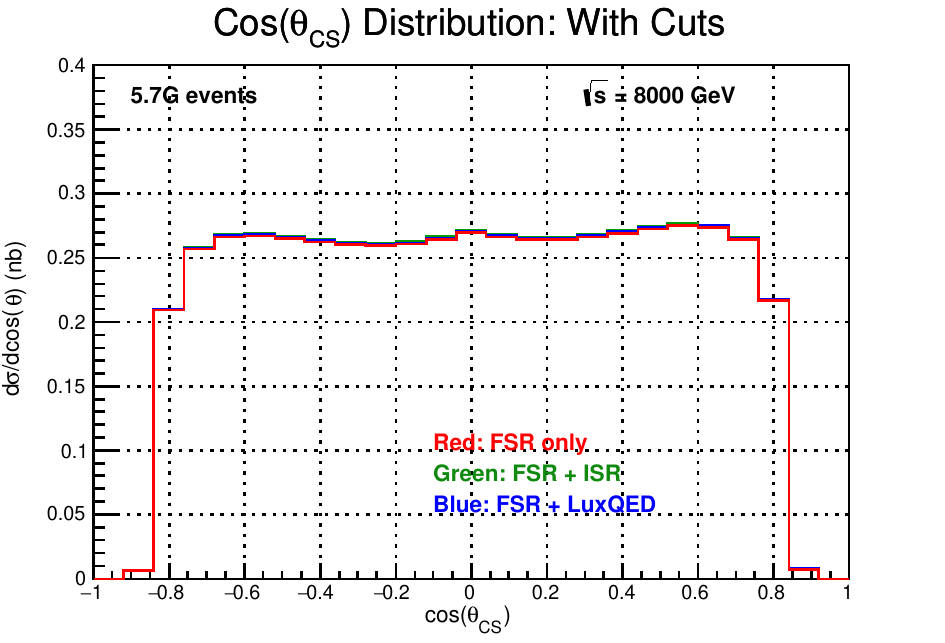}}
\put(3.2,.2){\includegraphics[width=3.2in,height=2.0in]{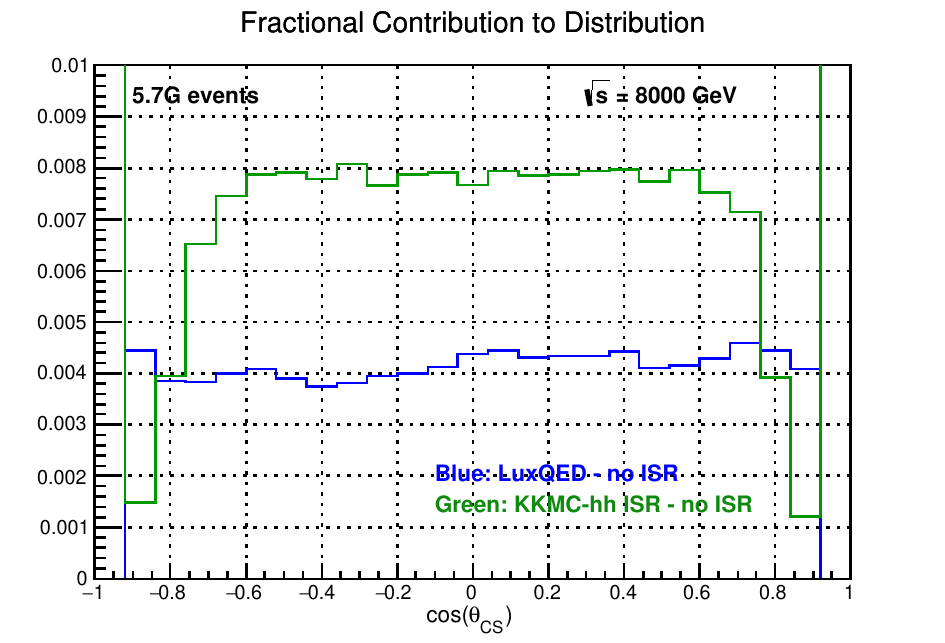}}
\end{picture}
\end{center}
\vspace{-10mm}
\caption{\baselineskip=11pt ISR contribution to the CS angle distribution: the two top plots are without lepton cuts (used for $A_4$ ), the two bottom plots are with lepton cuts (used for $A_{FB}$). LuxQED ISR + FSR is in blue(dark shade),  {\KK}MC-hh ISR+ FSR is in green (light shade), and FSR only, the baseline here, is in red (medium dark shade). In the two plots on the right,
the respective FSR only plot is subtracted from the  LuxQED ISR + FSR (blue) and {\KK}MC-hh ISR + FSR (green) plots.}
\label{fig2}
\end{figure}

We turn next to the ISR contributions to the distribution of the cosine of the CS angle. This is shown in Fig.~\ref{fig2}.
We see that ISR enters at the per mille level. It must be taken into account in precision studies of this process\footnote{As in LEP we expect the experimental statistical errors in angular distribution measurements to reflect the size of the experimental errors and we need that the theory errors stay at or below ~1/3 of the experimental errors so that the former do not significantly affect the latter. As on can see, for example, from Table 9 in Ref.~\cite{atlas-angcf-16} the experimental statistical error on $A_4$ in the 5 - 8 GeV $p_T$ bin is 1 per mille based on 20.3 fb$^{-1}$ of data. Since there remain ~ 143.6 fb$^{-1}$ of data to be analyzed, we have a budget for the theory error at or below 1 per mille/$(3\times \sqrt{163.9/20.3})\cong 0.00012$. All effects that are at or above this level have to be taken into account.} .\par
\begin{figure}[h]
\begin{center}
\setlength{\unitlength}{1in}
\begin{picture}(6.5,4.7)(0,0)
\put(0.2,2.5){\includegraphics[width=3.2in,height=2.0in]{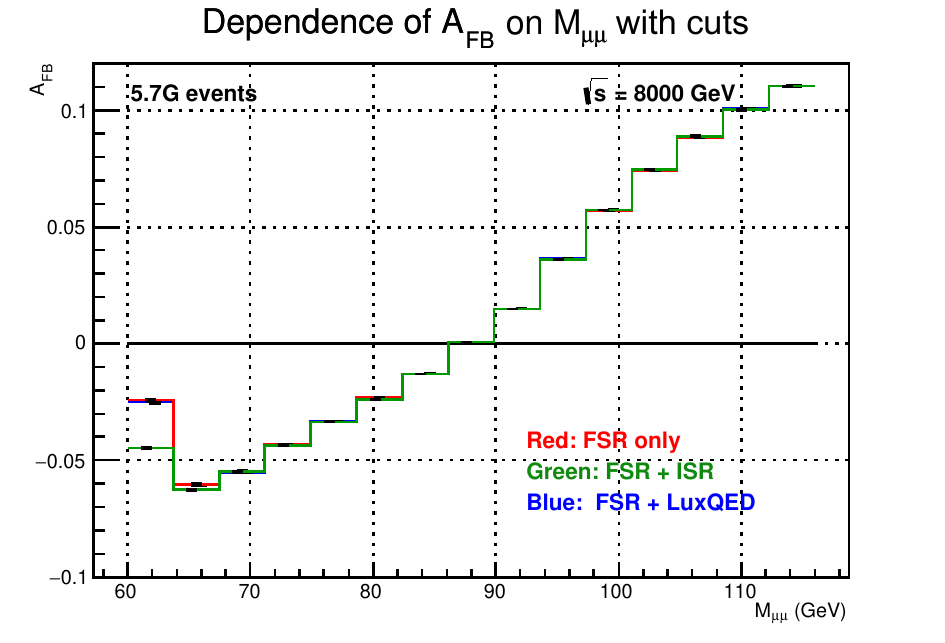}}
\put(3.2,2.5){\includegraphics[width=3.2in,height=2.0in]{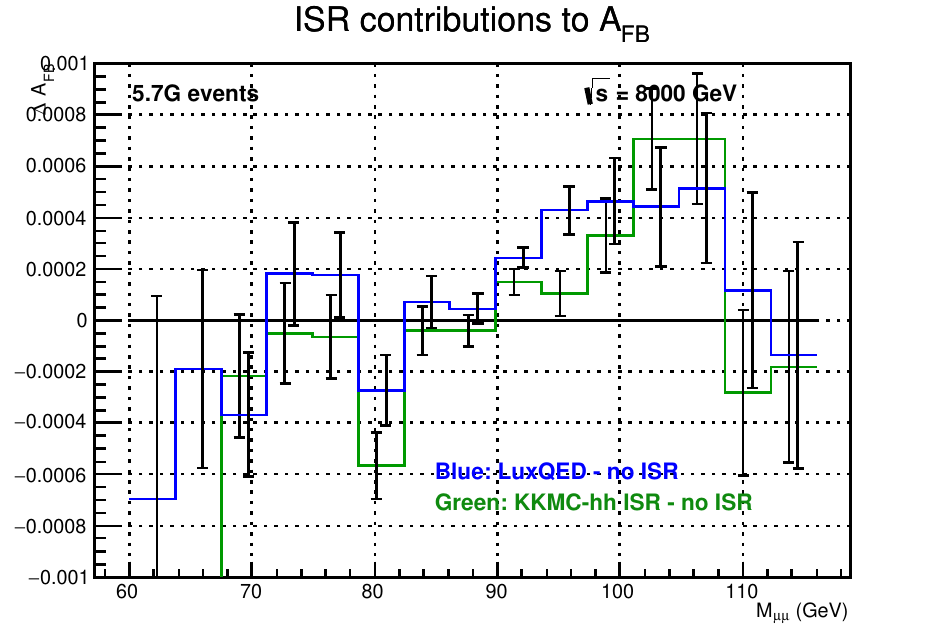}}
\put(0.2,.2){\includegraphics[width=3.2in,height=2.0in]{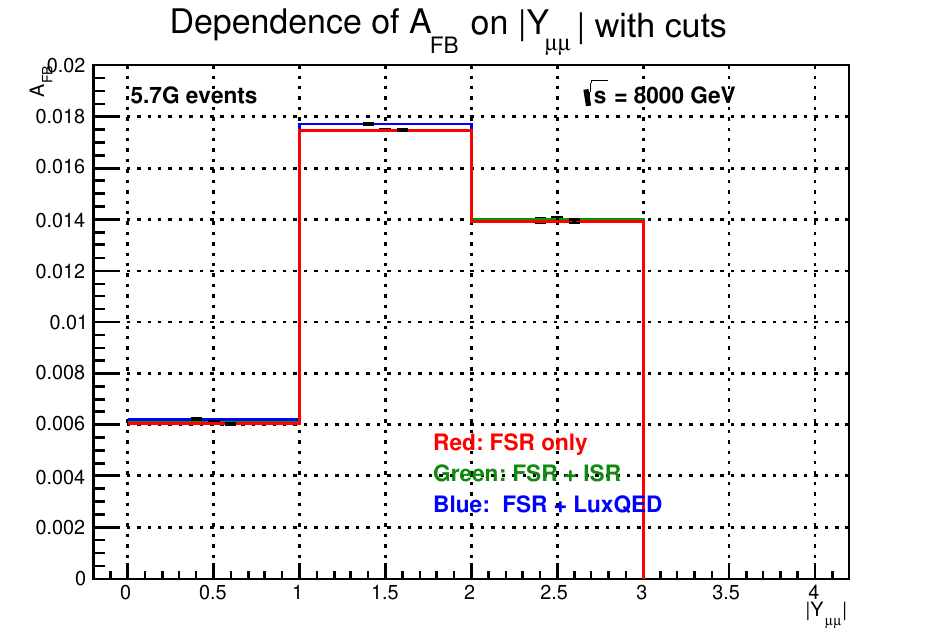}}
\put(3.2,.2){\includegraphics[width=3.2in,height=2.0in]{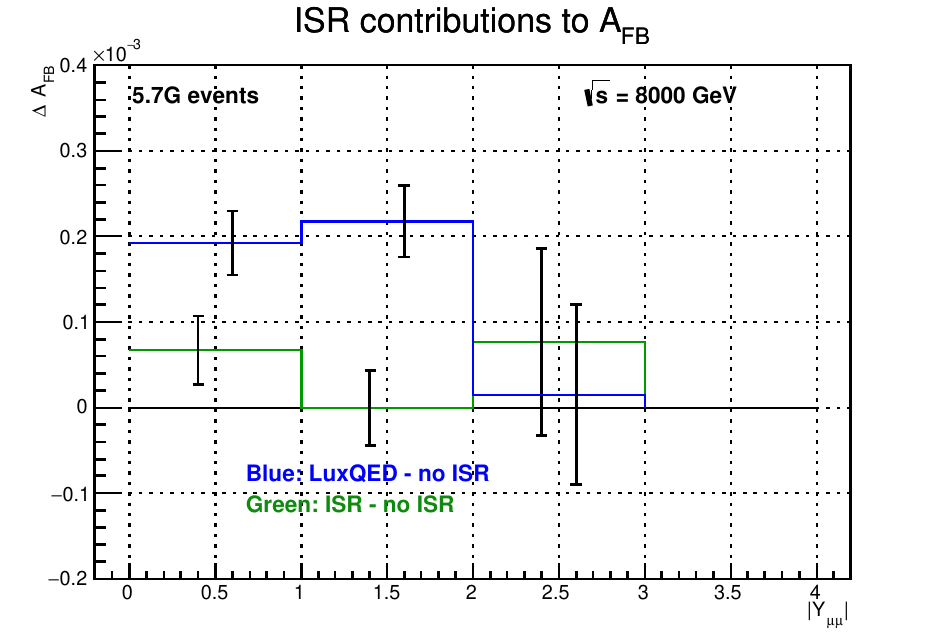}}
\end{picture}
\end{center}
\vspace{-10mm}
\caption{\baselineskip=11pt  ISR contributions to $A_{FB}$ with lepton cuts. Results are shown for FSR only, LuxQED ISR + FSR and \KK{MC}-hh ISR + FSR. The color scheme for the plots is the same as that in Fig.~\ref{fig2}. The plots on the right show the respective differences between the FSR only plot and the plots with ISR + FSR. Results are plotted as functions of $M_{\ell\ell}$
(top plots) and as functions of $|Y_{\ell\ell}|$.}
\label{fig3}
\end{figure}

To see how the effects in Fig.~\ref{fig2} affect the angular coefficients, we turn next to results for $A_{FB}$ as shown in Fig.~\ref{fig3}.
The ISR contribution to $A_{FB}$ is typically on the 
per mille level.  For most $M_{\ell\ell}$ of interest,  LuxQED and {\KK}MC-hh produce very similar ISR effects.
Integrating over$M_{\ell\ell}$  and binning in $|Y_{\ell\ell}|$, both LuxQED and {\KK}MC-hh give ISR contributions on the order of $10^{-4}$, with
the {\KK}MC-hh correction smaller at low rapidities.  It should be taken into account in precision studies of this process.\par

\begin{figure}[h]
\begin{center}
\setlength{\unitlength}{1in}
\begin{picture}(6.5,4.7)(0,0)
\put(0.2,2.5){\includegraphics[width=3.2in,height=2.0in]{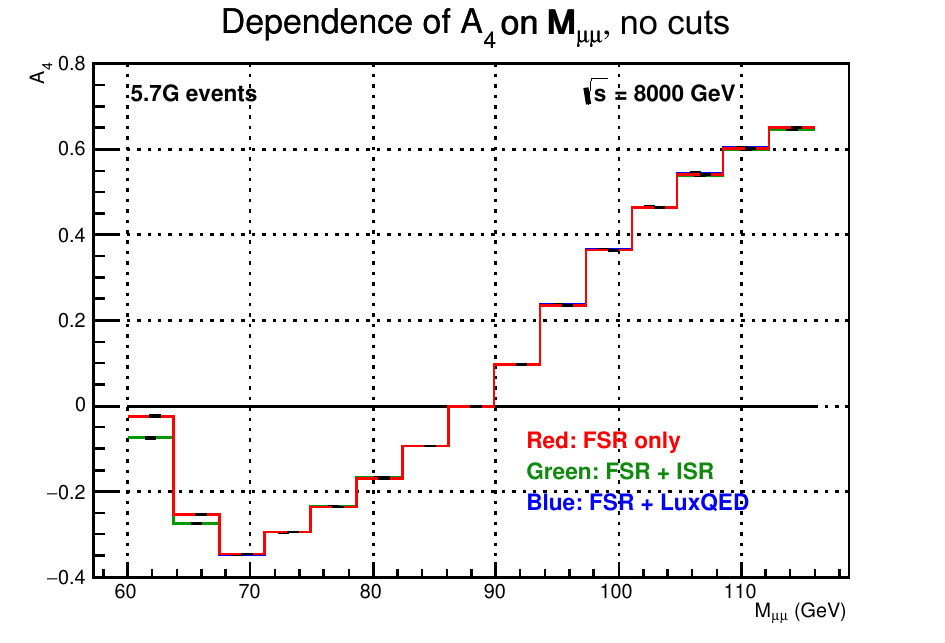}}
\put(3.2,2.5){\includegraphics[width=3.2in,height=2.0in]{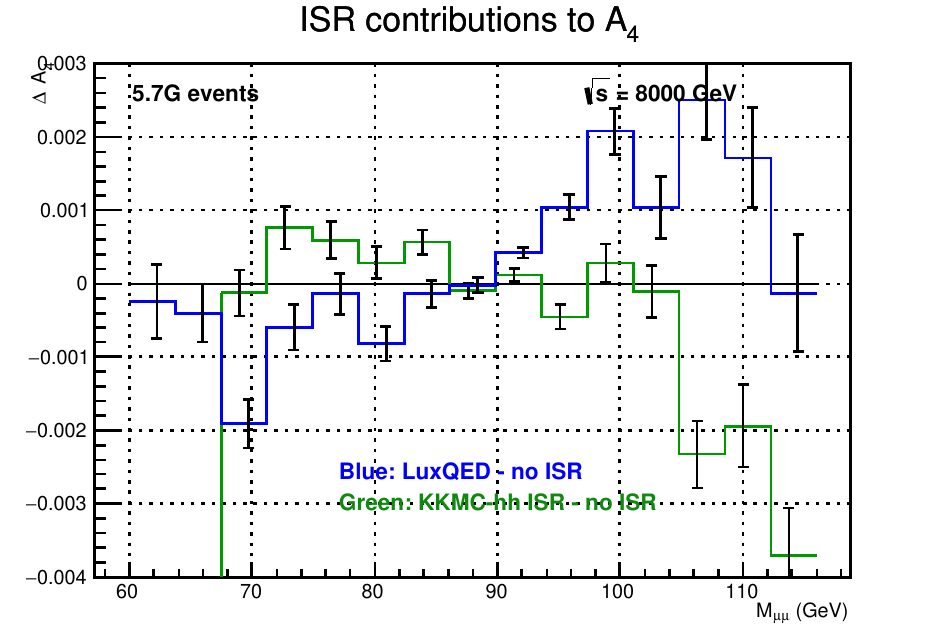}}
\put(0.2,.2){\includegraphics[width=3.2in,height=2.0in]{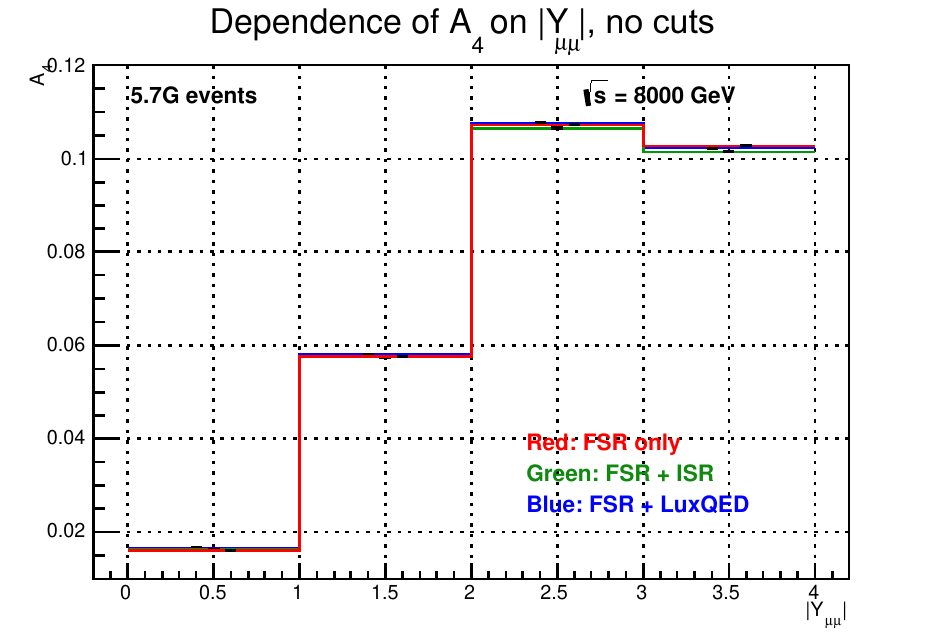}}
\put(3.2,.2){\includegraphics[width=3.2in,height=2.0in]{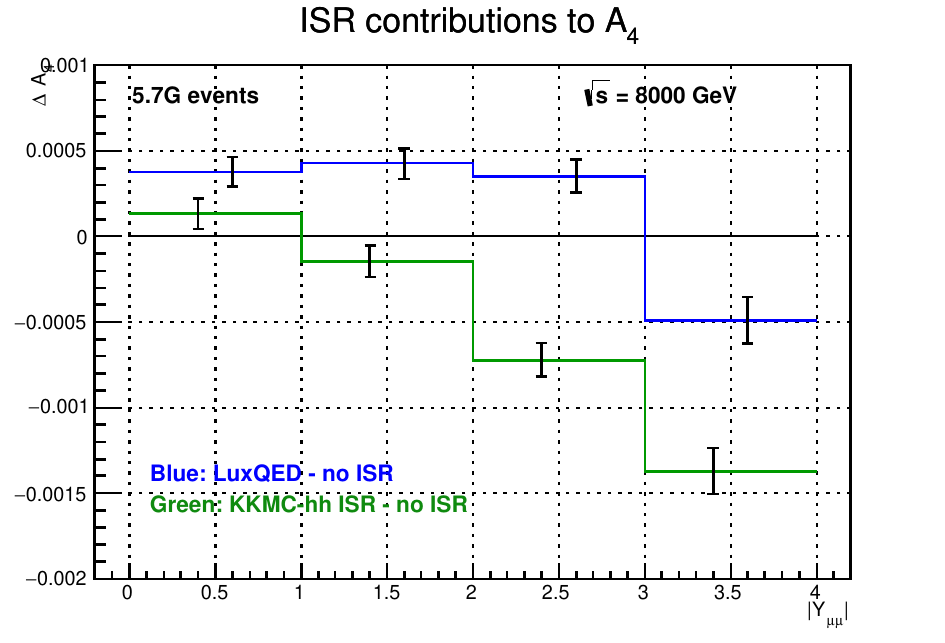}}
\end{picture}
\end{center}
\vspace{-10mm}
\caption{\baselineskip=11pt  ISR contributions to $A_4$ without lepton cuts, to remain consistent with the formula we use for it. Results are shown for FSR only, LuxQED ISR + FSR and {\KK}MC-hh ISR + FSR. The color scheme for the plots is the same as that in Fig.~\ref{fig2}. The plots on the right show the respective differences between the FSR only plot and the plots with ISR + FSR. Results are plotted as functions of $M_{\ell\ell}$
(top plots) and as functions of $|Y_{\ell\ell}|$ (bottom plots).}
\label{fig4}
\end{figure}

Continuing in this way, we show our results for the angular coefficient $A_4$ in Fig.~\ref{fig4}. 
The ISR contribution to  $A_4$ is typically on the order of $10^{-3}$, but differs in detail between LuxQED and {\KK}MC-hh.
When integrated over $M_{\ell\ell}$ and binned in $|Y_{\ell\ell}|$, the ISR contribution is a little smaller, and of order $10^{-4}$ for {\KK}MC-hh at low rapidities.
It should be taken into account in precision studies of this process.\par
We turn next to the initial-state-final-state interference (IFI) effects. We point-out that, due to IFI,
it is not possible to separate  unambiguously photon radiation into ISR and FSR. This complicates the interpretation of  $A_{FB}$ and $A_4$ unless IFI can be shown to be sufficiently small. Exponentiation at the amplitude level (CEEX), instead of the cross section level (EEX) facilitates the calculation of interference effects. This is one of the primary reasons CEEX was introduced, when effects at this level became relevant at LEP. IFI is implemented in CEEX by dividing the generated photons into partitions of ISR and FSR, and summing over all such partitions. In the following, we compare {\KK}MC-hh results with IFI turned on or off. The effect on angular variables is shown in terms of $M_{\ell\ell}$ and $|Y_{\ell\ell}|$ bins.\par
\begin{figure}[h]
\begin{center}
\setlength{\unitlength}{1in}
\begin{picture}(6.5,4.7)(0,0)
\put(0.2,2.5){\includegraphics[width=3.2in,height=2.0in]{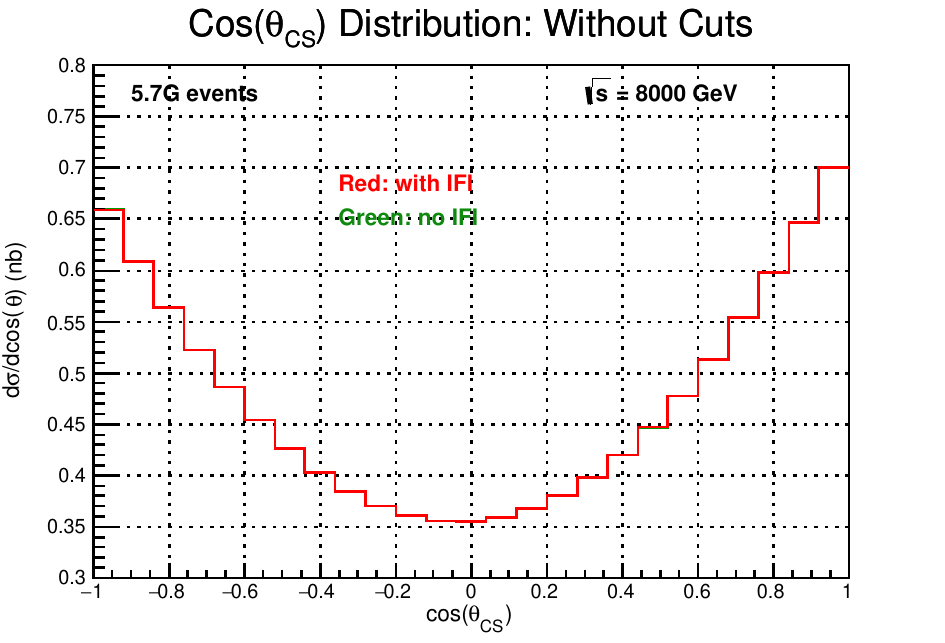}}
\put(3.2,2.5){\includegraphics[width=3.2in,height=2.0in]{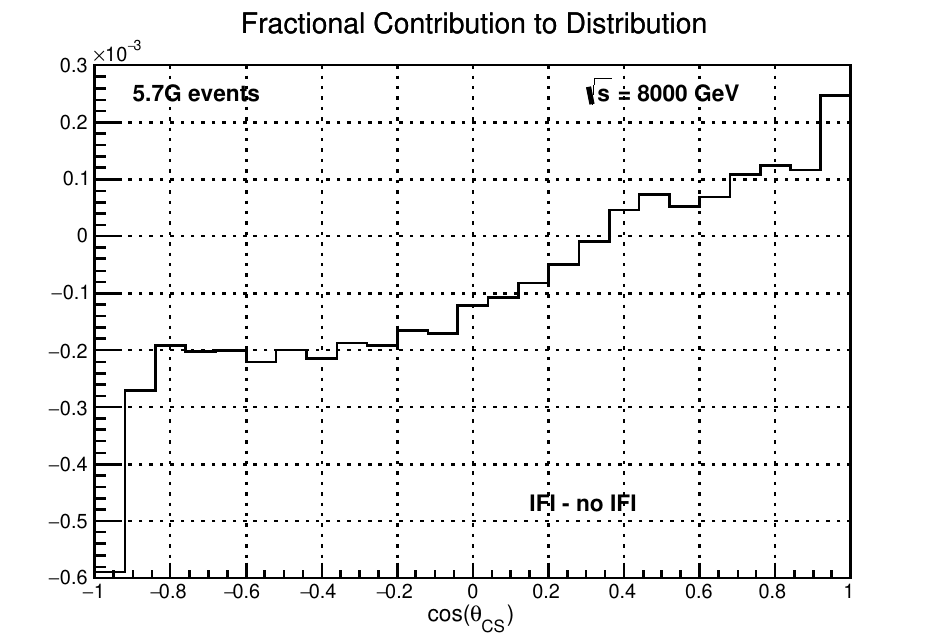}}
\put(0.2,.2){\includegraphics[width=3.2in,height=2.0in]{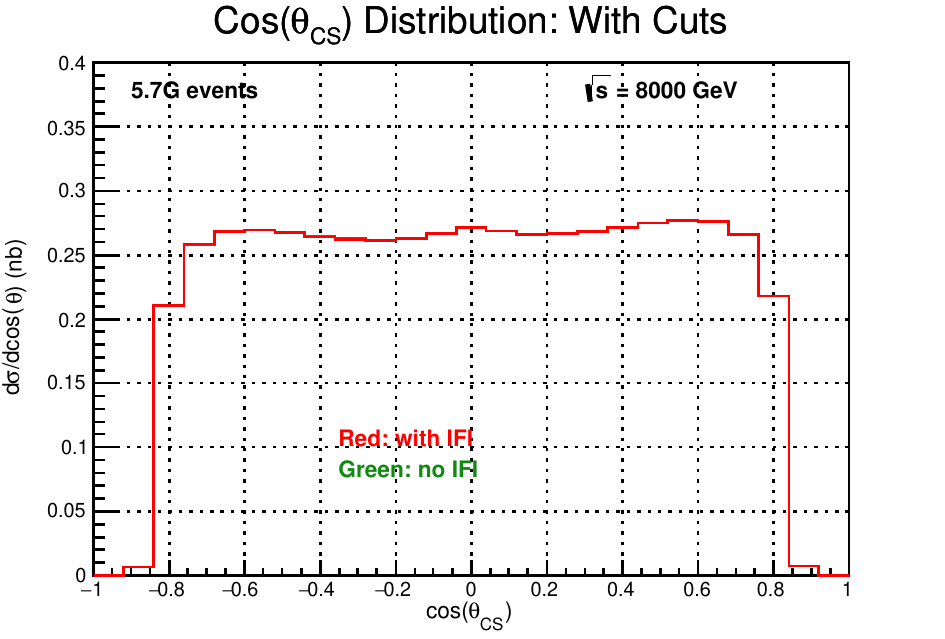}}
\put(3.2,.2){\includegraphics[width=3.2in,height=2.0in]{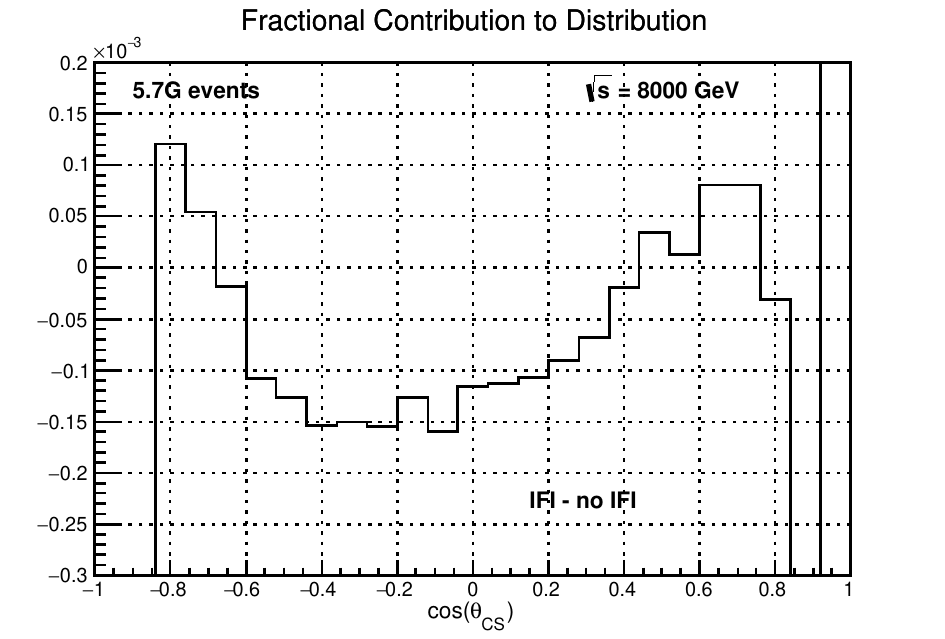}}
\end{picture}
\end{center}
\vspace{-10mm}
\caption{\baselineskip=11pt  IFI contribution to the distribution of $\cos\theta_{CS}$  without lepton cuts (for $A_4$) in the top plots and  with lepton cuts (for $A_{FB}$) in the bottom plots. Results are shown for {\KK}MC-hh ISR + FSR in red (medium dark shade) and for  \KK{MC}-hh ISR + FSR+IFI in green (light shade). The plots on the right show the respective fractional IFI contribution to the distributions on the left.}
\label{fig5}
\end{figure}

We study first the effects of IFI on the distribution of the cosine of the CS angle with and without lepton cuts in Fig.~\ref{fig5}.
Both with and without the lepton cuts, there are IFI effects at the $10^{-4}$ level, but with very different dependencies on $\cos\theta_{CS}$. To be on the safe side, precision studies should take these effects into account.\par
\begin{figure}[h]
\begin{center}
\setlength{\unitlength}{1in}
\begin{picture}(6.5,4.7)(0,0)
\put(0.2,2.5){\includegraphics[width=3.2in,height=2.0in]{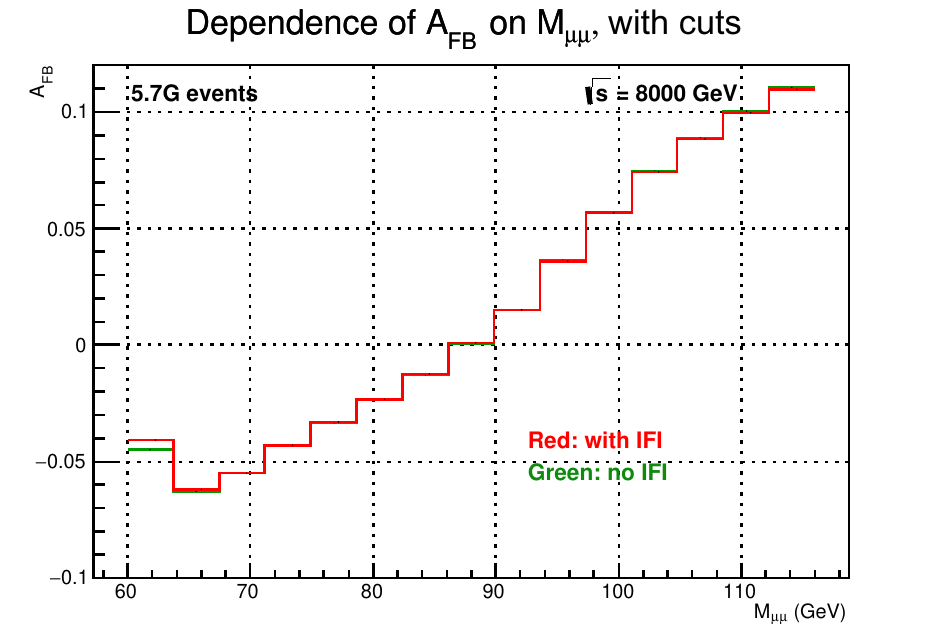}}
\put(3.2,2.5){\includegraphics[width=3.2in,height=2.0in]{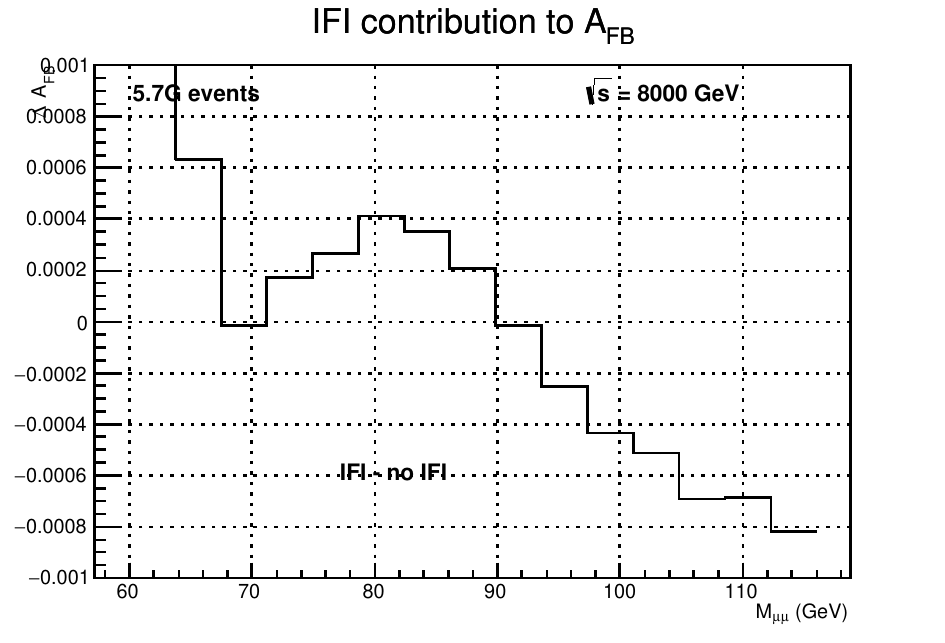}}
\put(0.2,.2){\includegraphics[width=3.2in,height=2.0in]{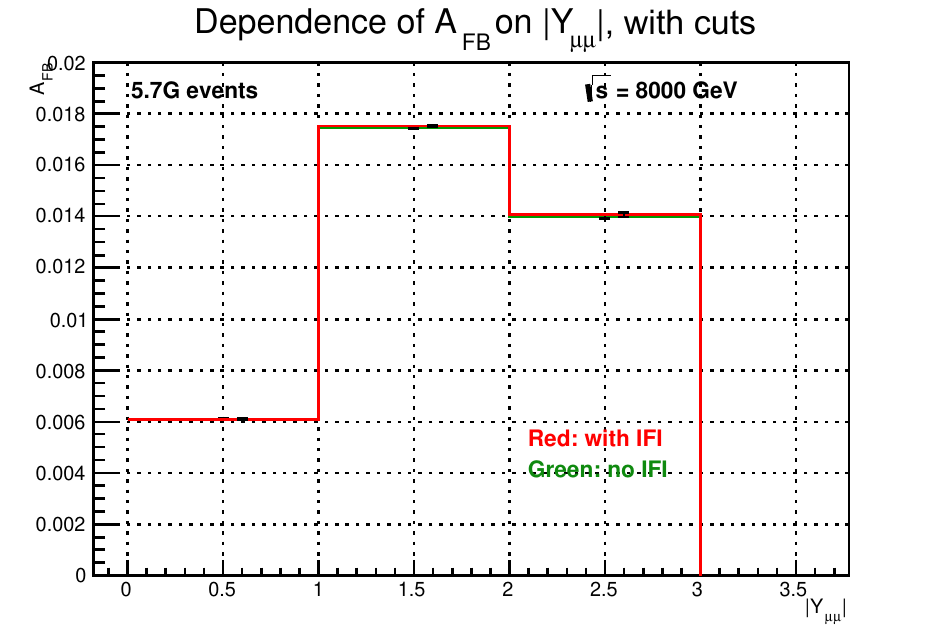}}
\put(3.2,.2){\includegraphics[width=3.2in,height=2.0in]{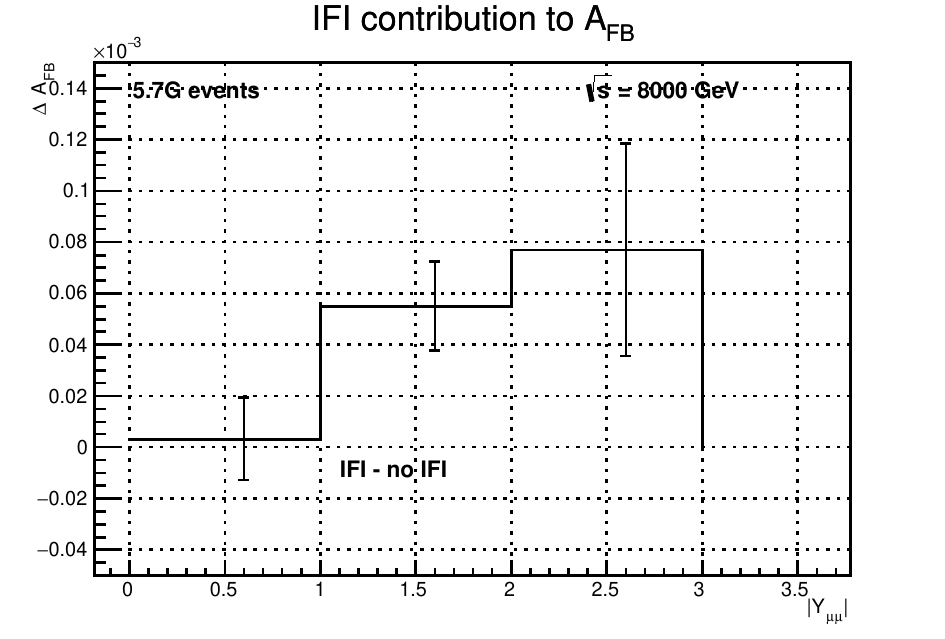}}
\end{picture}
\end{center}
\vspace{-10mm}
\caption{\baselineskip=11pt  IFI contribution to $A_{FB}$ with lepton cuts. Results are shown for {\KK}MC-hh ISR + FSR in red (medium dark shade) and for  \KK{MC}-hh ISR + FSR+IFI in green (light shade). The plots on the right show the respective IFI contribution to the distributions on the left. Results are plotted as functions of $M_{\ell\ell}$
(top plots) and as functions of $|Y_{\ell\ell}|$ (bottom plots).}
\label{fig6}
\end{figure}

Focusing next of the angular observables, we show the IFI effects on $A_{FB}$ in Fig.~\ref{fig6}.
The IFI contribution to $A_{FB}$ is generally less than $10^{-3}$. When integrated over $M_{\ell\ell}$, the IFI contribution is typically less than $10^{-4}$, and much less for small rapidities.
In general, precision studies should take this contribution into account.\par
\begin{figure}[h]
\begin{center}
\setlength{\unitlength}{1in}
\begin{picture}(6.5,4.7)(0,0)
\put(0.2,2.5){\includegraphics[width=3.2in,height=2.0in]{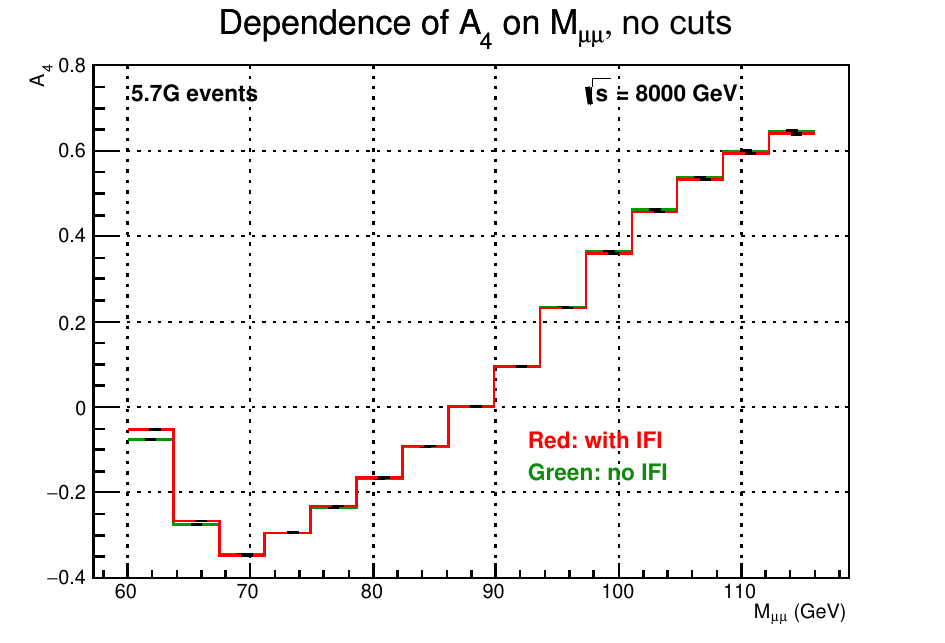}}
\put(3.2,2.5){\includegraphics[width=3.2in,height=2.0in]{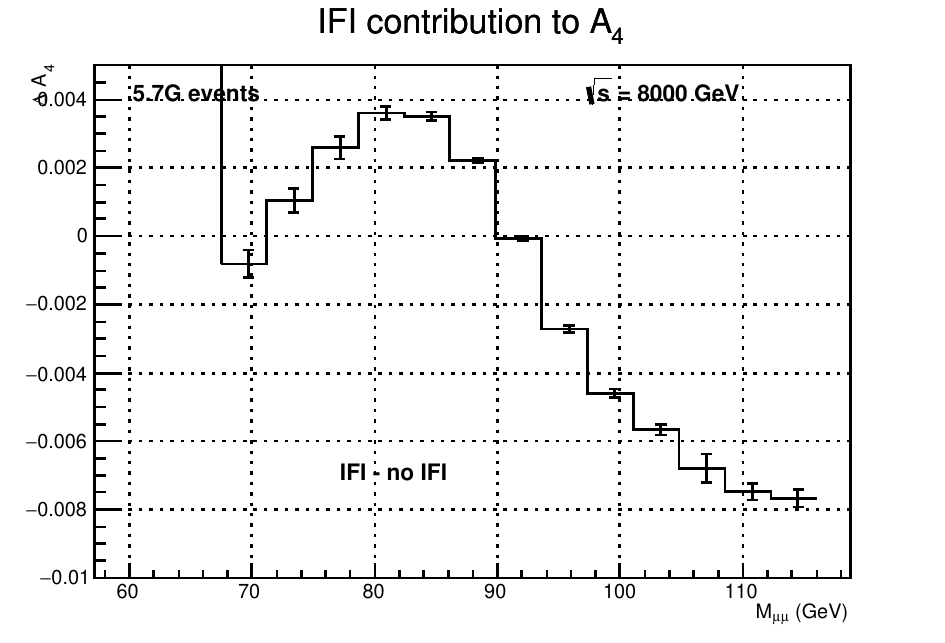}}
\put(0.2,.2){\includegraphics[width=3.2in,height=2.0in]{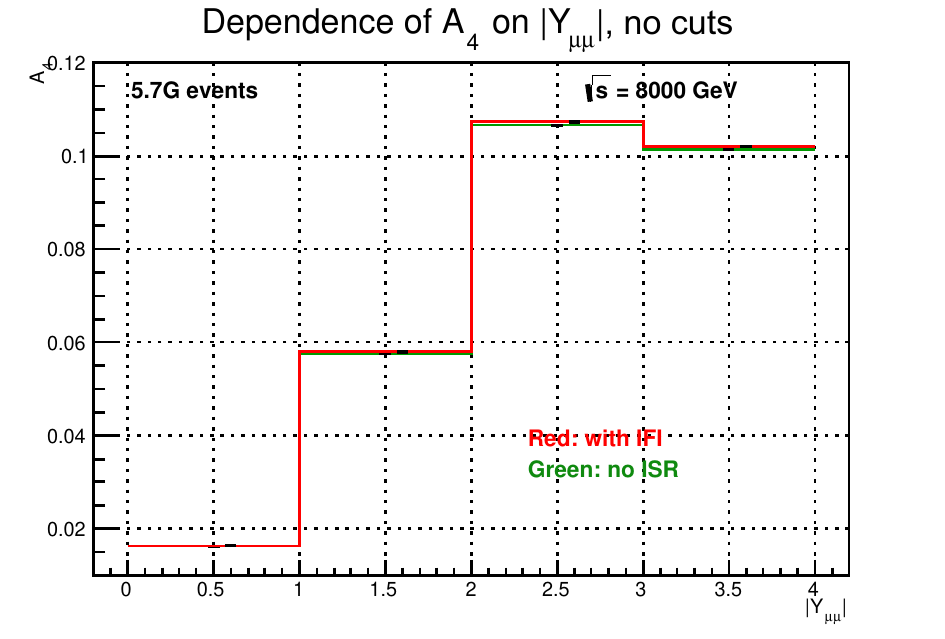}}
\put(3.2,.2){\includegraphics[width=3.2in,height=2.0in]{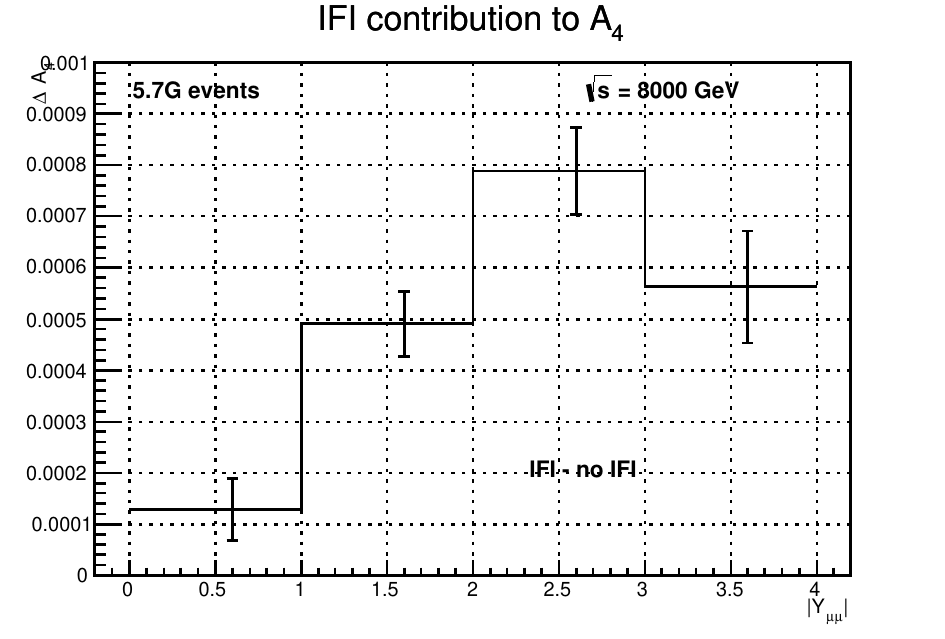}}
\end{picture}
\end{center}
\vspace{-10mm}
\caption{\baselineskip=11pt  IFI contribution to $A_{4}$ without lepton cuts. Results are shown for {\KK}MC-hh ISR + FSR in red (medium dark shade) and for  \KK{MC}-hh ISR + FSR+IFI in green (light shade). The plots on the right show the respective IFI contribution to the distributions on the left. Results are plotted as functions of $M_{\ell\ell}$
(top plots) and as functions of $|Y_{\ell\ell}|$ (bottom plots).}
\label{fig7}
\end{figure}

Similarly, we show the IFI effects on $A_{4}$ in Fig.~\ref{fig7}.
The IFI contribution to $A_{4}$ is generally less than $10^{-2}$ but depends on $M_{\ell\ell}$. When integrated over $M_{\ell\ell}$, the IFI contribution is generally less than $10^{-3}$, and very small for some rapidities.  Precision studies should take this contribution into account.\par
We turn now to the effect of the parton shower on the previous results. We use the built-in Herwg 6.521~\cite{HERWIG} shower but we stress that, due to the LHE format in {\KK}MC-hh, in principle any shower compatible with that format can be used. In Table ~\ref{tab-2}, we show the numerical effect of the shower on  results for the cross section, $A_{FB}$, and $A_4$ from {\KK}MC-hh with the ISR on and the non-QED NNPDF 3.1 NLO PDF. The results are determined from a sample of $1.1\times 10^9$ events at 8 TeV.
\begin{table}[h!]
\caption{\text{Showered Numerical Results: $\sigma,\; A_{FB},\; A_4$}}
\centering
\scalebox{.85}{
\begin{tabular}{|c|c|c|c|}
\hline
&{\small Without Shower}&{\small With Shower}&{\small \% Difference}\\
\hline
{\small\text{Uncut} $\sigma (\text{pb})$ }&{\small 944.91(2)}&{\small 938.44(4)}&{\small -0.684(7)\%} \\
{\small\text{Cut} $\sigma (\text{pb})$}&{\small 442.33(1)}&{\small 412.54(3)}&{\small -6.7307\%}\\
\hline
&{\small without Shower}&{\small With Shower}&{\small  Difference}\\
\hline
{\small$A_{FB}$}&{\small 0.01132(2)}&{\small 0.01211(5)}&{\small 0.00109(5)}\\
{\small$A_4$}&{\small 0.06102(8)}&{\small 0.06052(8)}&{\small -0.00050(8)}\\
\hline
\end{tabular}}
\label{tab-2}
\end{table}
The results are shown for both the cut and uncut cases.
For the uncut cases, we see effects at the \% level. For the cut cases, we see effects at the 7 - 8 \% level. We note that, as the shower does not affect the overall normalization, we expect smaller effects from the shower in the uncut scenarios and larger effects in the cut cases in which the available phase space is much more restricted. Our results support this expectation. Precision studies should take these effects into account.\par
Considering the IFI contributions to the cross section with and without cuts, we show the effects of the shower in Table ~\ref{tab-3}.
\begin{table}[h] 
\caption{\small\text{Showered Results: IFI Contributions to} $\sigma$}
\centering
\scalebox{.95}{
\begin{tabular}{|c|c|c|c|}
\hline
{\small Uncut $\sigma$}&{\small No IFI (pb)}&{\small With IFI (pb)}&{\small \% Difference}\\
\hline
{\small\text{No Shower}}&{\small 944.99(2)}&{\small 944.91(2)}&{\small -0.0089(4)\%} \\
{\small\text{Shower}}&{\small 938.46(4)}&{\small 938.44(4)}&{\small -0.002(1)\%}\\
{\small\text{Difference}}&{\small -0.691(5)\%}&{\small -0.684(5)\%}&{\small 0.007(1)\%}\\
\hline
{\small Cut $\sigma$}&{\small No IFI (pb)}&{\small With IFI (pb)}&{\small \% Difference}\\
\hline
{\small\text{No Shower}}&{\small 442.36(1)}&{\small 442.33(1)}&{\small -0.0070(5)\%} \\
{\small\text{Shower}}&{\small 412.54(3)}&{\small 412.56(3)}&{\small -0.004(2)\%}\\
{\small\text{Difference}}&{\small -6.741(7)\%}&{\small -6.730(7)\%}&{\small 0.003(2)\%}\\
\hline
\end{tabular}}
\label{tab-3}
\end{table}
In each case, the IFI contribution is significantly smaller with the shower on.\par
The comparisons between the showered and unshowered results for the IFI contributions to $A_{FB}$ and to $A_4$ are shown in Table ~\ref{tab-4}.
\begin{table}[h!]
\caption{\text{Showered Results: IFI Contributions to} $A_{FB}$ \text{and to} $A_4$ }
\centering
\scalebox{.95}{
\begin{tabular}{|c|c|c|c|}
\hline
{\small $A_{FB}$}&{\small No IFI (pb)}&{\small With IFI (pb)}&{\small Difference}\\
\hline
{\small\text{No Shower}}&{\small 0.01129(2)}&{\small 0.01132(2)}&{\small $(2.9\pm1.1)\times10^{-5}$} \\
{\small\text{Shower}}&{\small 0.01235(5)}&{\small 0.01241(5)}&{\small $(5.8\pm 2.6)\times10^{-5}$}\\
{\small\text{Difference}}&{\small 0.00106(5)}&{\small 0.00109(5)}&{\small $(2.9\pm 2.8)\times10^{-5}$}\\
\hline
{\small $A_4$}&{\small No IFI (pb)}&{\small With IFI (pb)}&{\small  Difference}\\
\hline
{\small\text{No Shower}}&{\small 0.06057(3)}&{\small 0.06102(3)}&{\small $(4.5\pm 0.3)\times10^{-4}$} \\
{\small\text{Shower}}&{\small 0.06003(8)}&{\small 0.06052(8)}&{\small $(4.9\pm 0.8)\times10^{-4}$}\\
{\small\text{Difference}}&{\small -0.00055(8)}&{\small -0.00050(8)}&{\small $(4.3\pm 8.5)\times10^{-5}$}\\
\hline
\end{tabular}}
\label{tab-4}
\end{table}
The effect of the shower on the IFI contribution is statistically insignificant for $A_4$ and is barely significant, of order $10^{-5}$, for $A_{FB}$.\par
\begin{figure}[h]
\begin{center}
\setlength{\unitlength}{1in}
\begin{picture}(6.5,4.7)(0,0)
\put(0.2,2.5){\includegraphics[width=3.2in,height=2.0in]{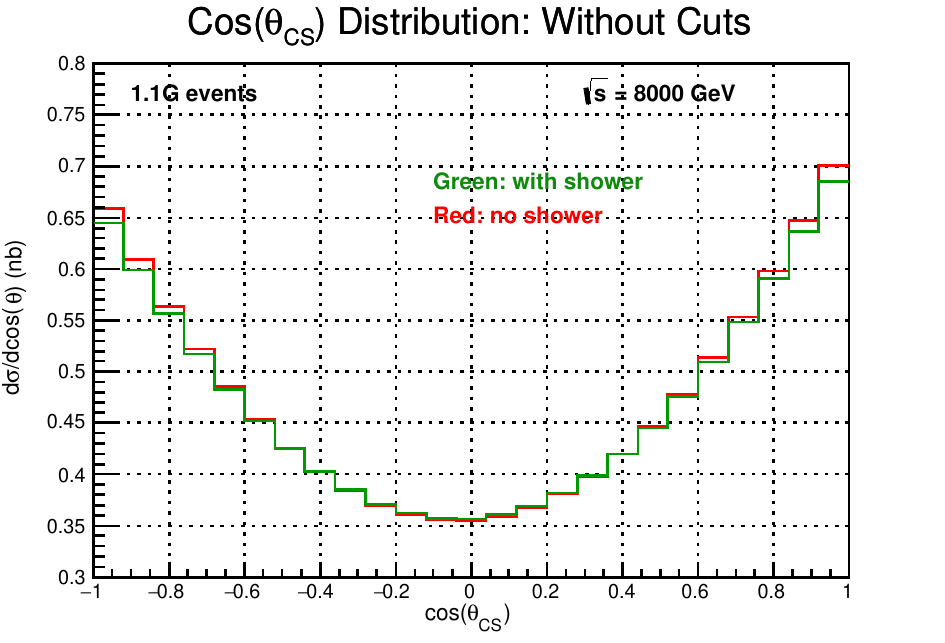}}
\put(3.2,2.5){\includegraphics[width=3.2in,height=2.0in]{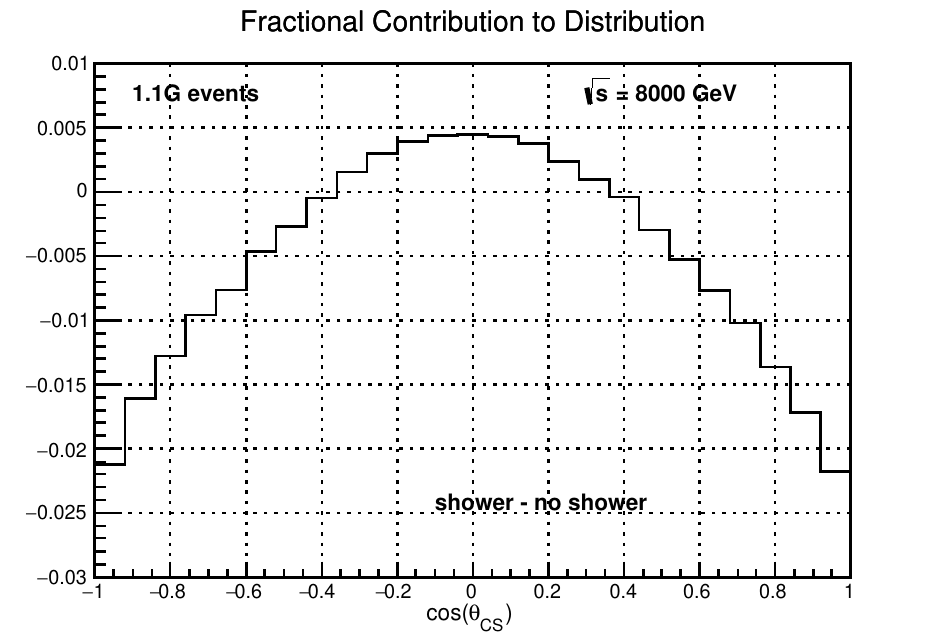}}
\put(0.2,.2){\includegraphics[width=3.2in,height=2.0in]{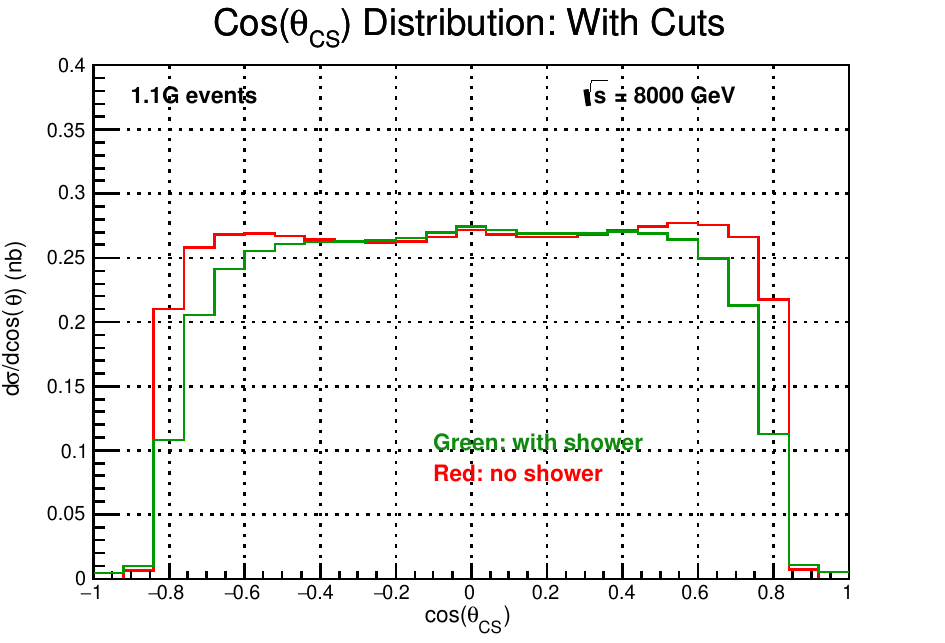}}
\put(3.2,.2){\includegraphics[width=3.2in,height=2.0in]{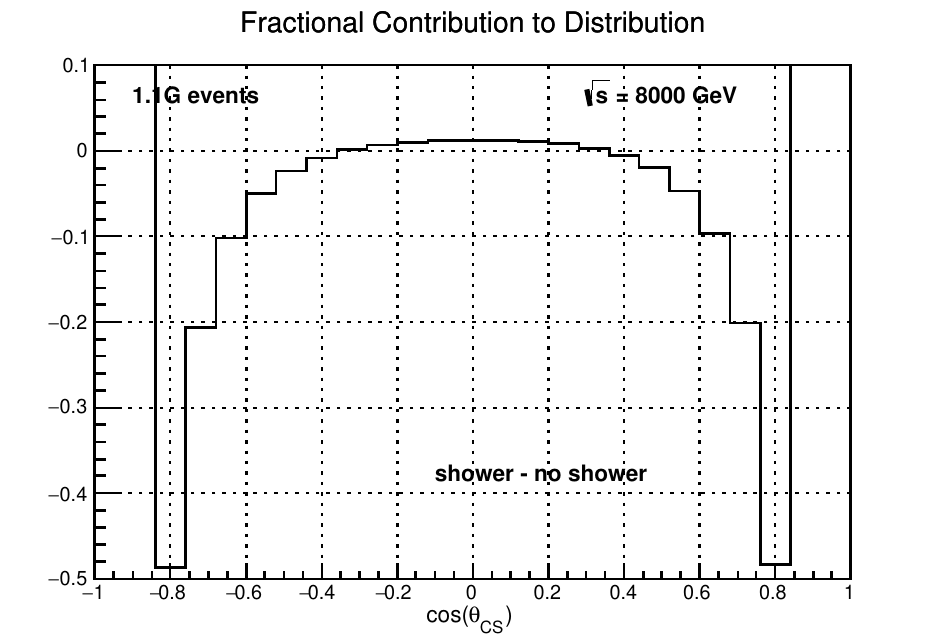}}
\end{picture}
\end{center}
\vspace{-10mm}
\caption{\baselineskip=11pt Showered contributions to the CS angle distribution: the two top plots are without lepton cuts (used for $A_4$ ), the two bottom plots are with lepton cuts (used for $A_{FB}$). Showered {\KK}MC-hh ISR+ FSR results are shown in green (light shade), and unshowered {\KK}MC-hh ISR+ FSR results are shown in red (medium dark shade). In the two plots on the right,
the respective fractional contributions of the shower effects to the distributions are shown.}
\label{fig8}
\end{figure}

We turn next to the effects of the shower on the angular distribution plotted as a function of $\cos(\theta_{CS})$ which we exhibit in Fig.~\ref{fig8}.
We see that the shower effects enter at the per cent level without cuts and enter at the 10-20\% level with cuts.\par
In Fig.~9 we show the effect of the shower on the IFI contribution, calculated with {\KK}MC-hh, to the uncut and cut CS angle distributions.
\begin{figure}[hb!]
\begin{center}
\setlength{\unitlength}{1in}
\begin{picture}(6.5,3.0)
\put(0,0.5){\includegraphics[width=3.2in,height=2.6in]{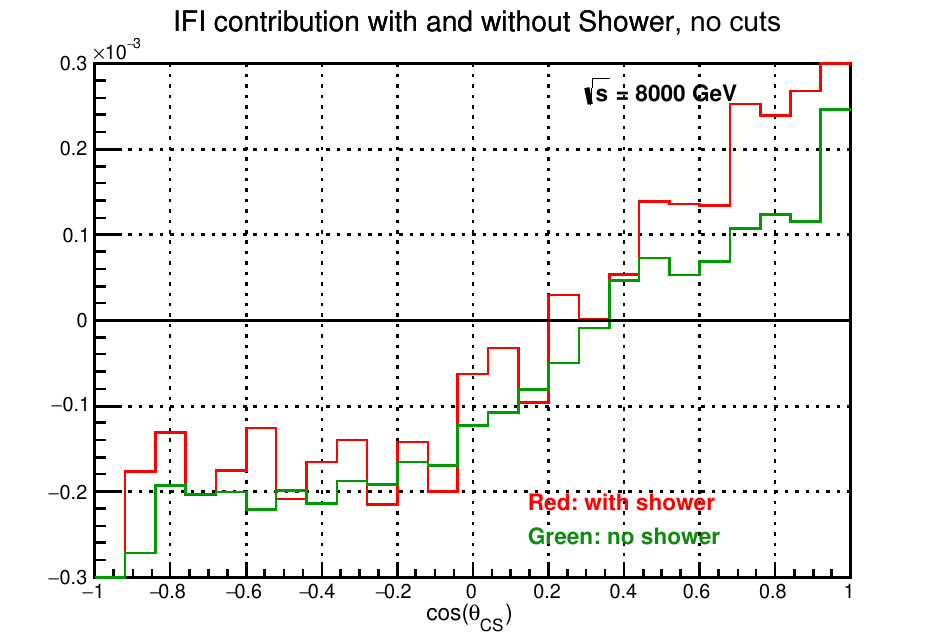}}
\put(3.0,0.5){\includegraphics[width=3.2in,height=2.6in]{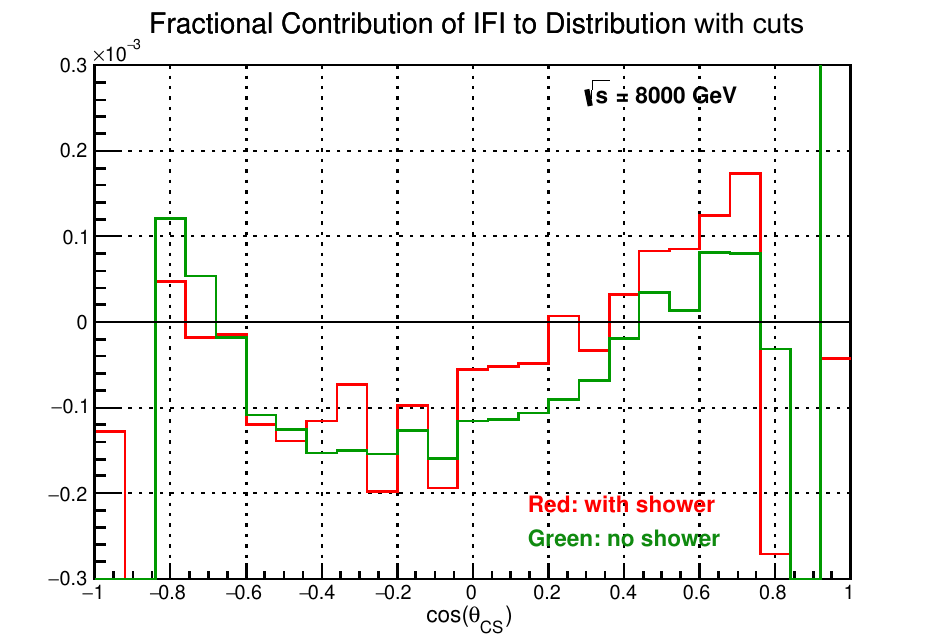}}
\end{picture}
\vspace{-0.75in}
\caption{Dependence of the Collins-Soper angular distribution on initial-final interference, without lepton cuts (left) and with them (right). Unshowered  results are shown in green (light shade), and showered results are shown in red (medium dark shade). }
\end{center}
\label{fig9}
\end{figure}
The IFI effect is angle-dependent and is at the level of a fraction of a per mille and the shower produces an angle-dependent modulation which still leaves the effect at the fractional per mille level.\par
\begin{figure}[h]
\begin{center}
\setlength{\unitlength}{1in}
\begin{picture}(6.5,4.7)(0,0)
\put(0.2,2.5){\includegraphics[width=3.2in,height=2.0in]{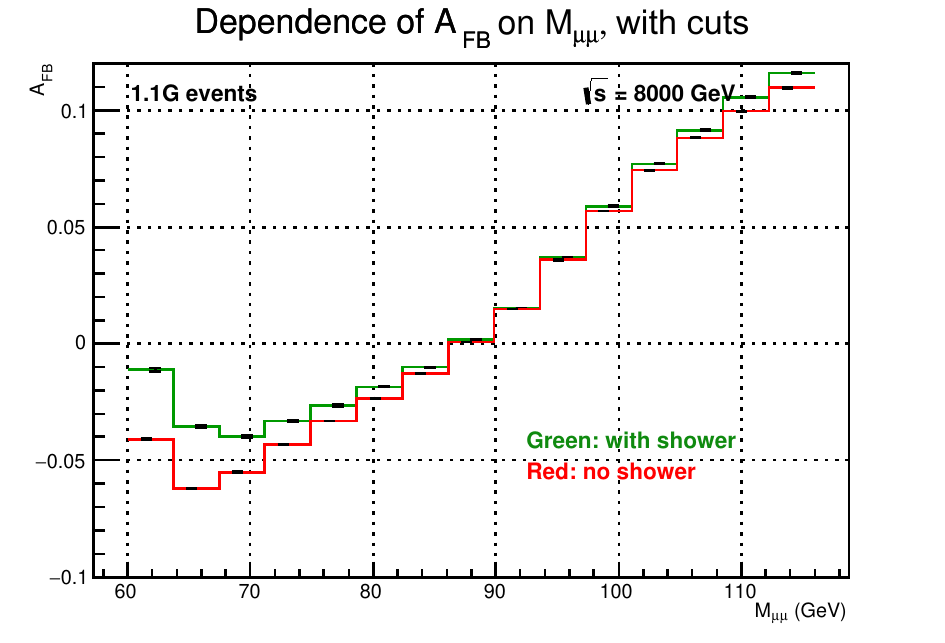}}
\put(3.2,2.5){\includegraphics[width=3.2in,height=2.0in]{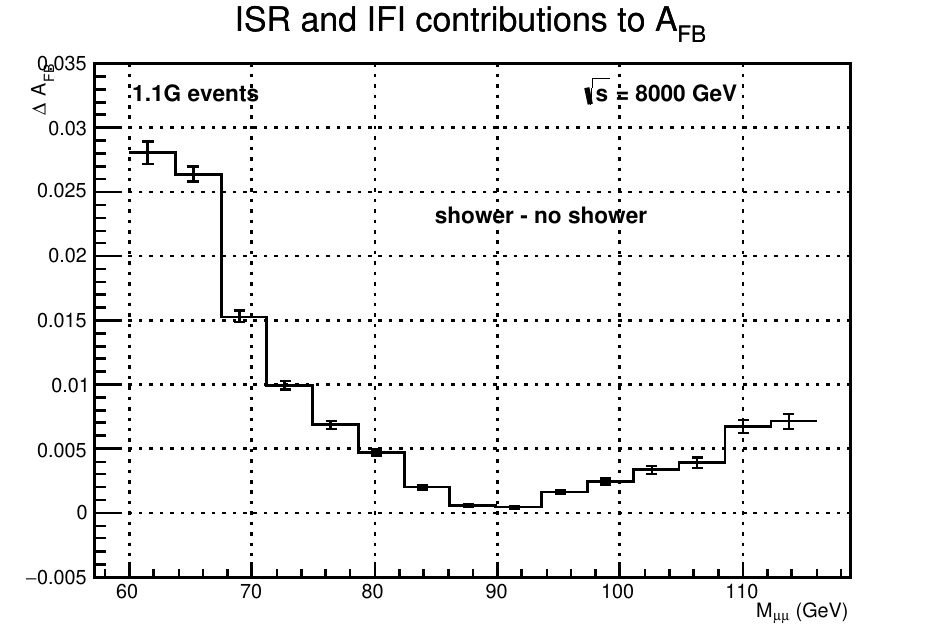}}
\put(0.2,.2){\includegraphics[width=3.2in,height=2.0in]{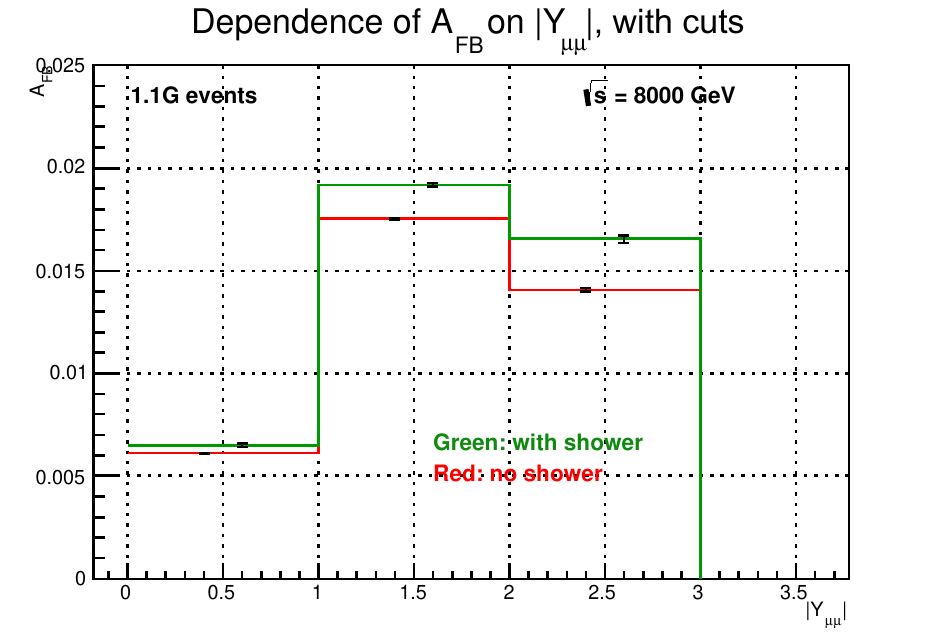}}
\put(3.2,.2){\includegraphics[width=3.2in,height=2.0in]{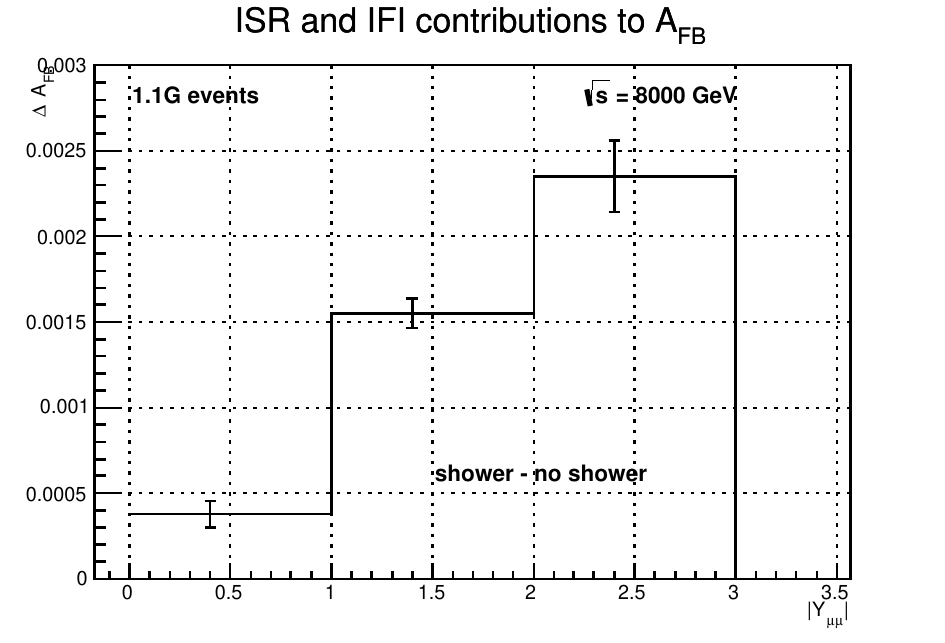}}
\end{picture}
\end{center}
\vspace{-10mm}
\caption{\baselineskip=11pt Showered contributions to  $A_{FB}$: the two top plots show the shower effects as a function of $M_{\ell\ell}$, the two bottom plots show the shower effects as a function of $Y_{\ell\ell}$. Showered {\KK}MC-hh ISR+ FSR+IFI results are shown in green (light shade), and unshowered {\KK}MC-hh ISR+ FSR+IFI results are shown in red (medium dark shade). In the two plots on the right,
the respective differences between the showered and unshowered distributions are shown.}
\label{fig10}
\end{figure}

In Fig.~\ref{fig10} we show the effects of the shower on $A_{FB}$ as a function of $M_{\ell\ell}$ and as a function of $Y_{\ell\ell}$.
The effect of the shower on 
increases for $M_{\ell\ell}$ away from $M_Z$ where  $A_{FB}$ is suppressed. The effect of the shower on $A_{FB}$ increases for
larger rapidities $Y_{\ell\ell}$. It is well-known that precision studies should take these effects into account.\par
\begin{figure}[h]
\begin{center}
\setlength{\unitlength}{1in}
\begin{picture}(6.5,3.0)(0,0)
\put(0,0.5){\includegraphics[width=3.2in,height=2.6in]{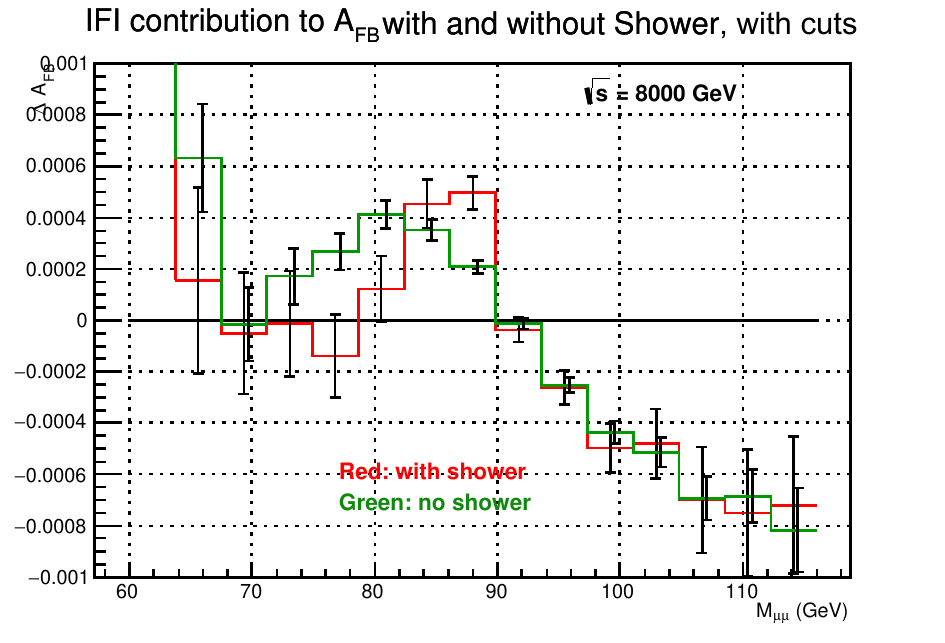}}
\put(3.0,0.5){\includegraphics[width=3.2in,height=2.6in]{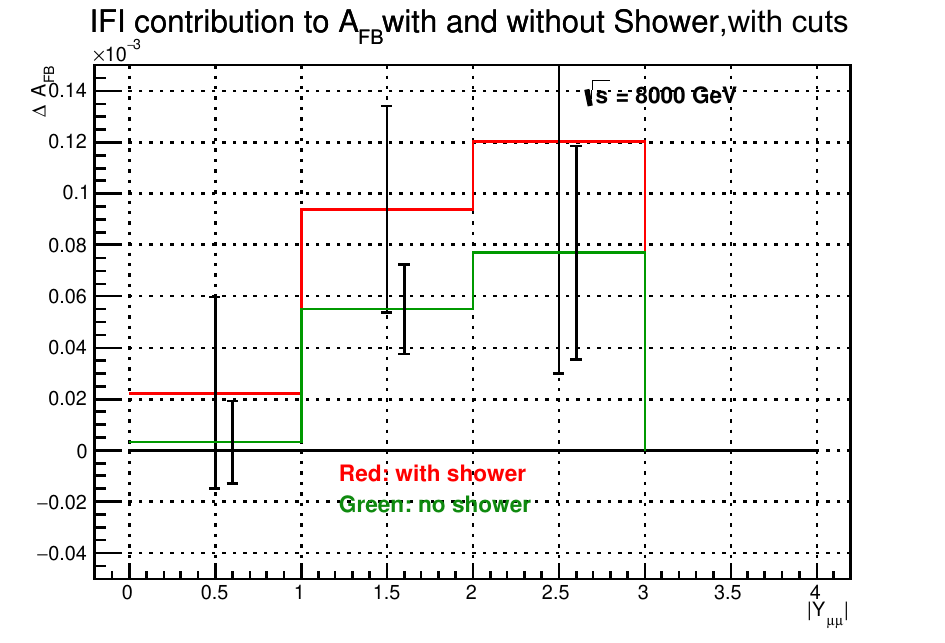}}
\end{picture}
\end{center}
\vspace{-10mm}
\caption{\baselineskip=11pt Shower effects on the IFI contribution to $A_{FB}$: the plots on the left illustrate the shower effects as a function of $M_{\ell\ell}$, the plots on the right illustrate the shower effects as a function of $Y_{\ell\ell}$. The results show the showered and unshowered differences between {\KK}MC-hh ISR+ FSR+IFI results and the corresponding {\KK}MC-hh ISR+ FSR results. The unshowered results are are in green (light shade), the showered results are in red (medium dark shade).}
\label{fig11}
\end{figure}

Similarly, in Fig.~\ref{fig11}, we show the shower effects on the IFI contribution to $A_{FB}$ as a function of $M_{\ell\ell}$ and as a function of $Y_{\ell\ell}$. 
The shower gives a mild modulation of the IFI effect as a function of $M_{\ell\ell}$; for the dependence on $Y_{\ell\ell}$, modulation is within the statistical errors.\par
\begin{figure}[h]
\begin{center}
\setlength{\unitlength}{1in}
\begin{picture}(6.5,4.7)(0,0)
\put(0.2,2.5){\includegraphics[width=3.2in,height=2.0in]{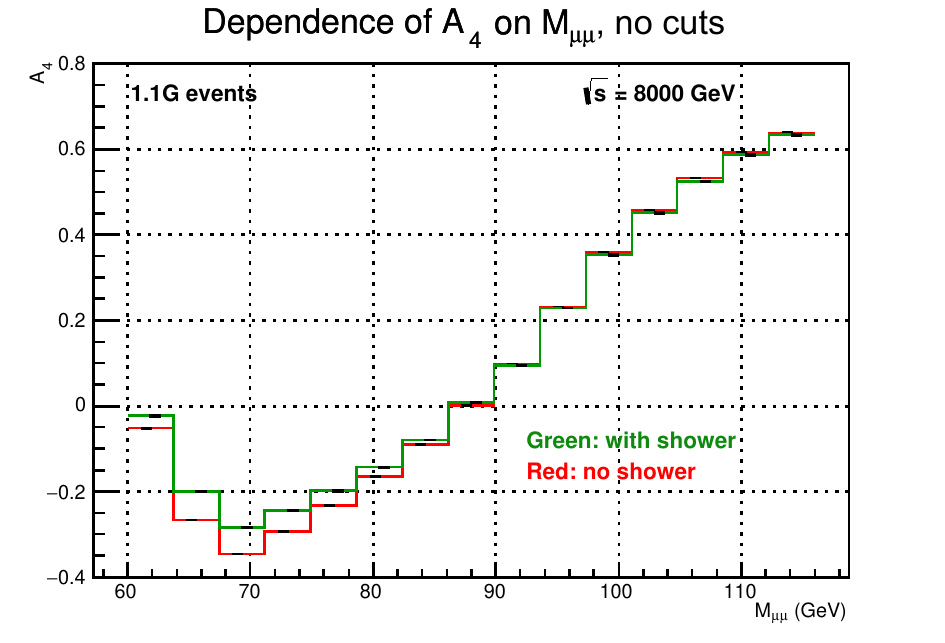}}
\put(3.2,2.5){\includegraphics[width=3.2in,height=2.0in]{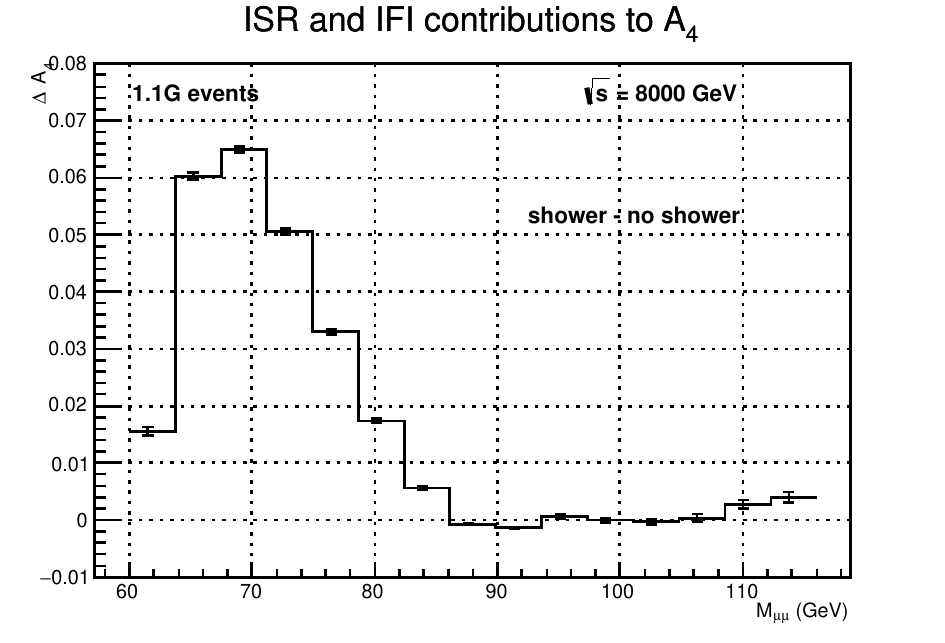}}
\put(0.2,.2){\includegraphics[width=3.2in,height=2.0in]{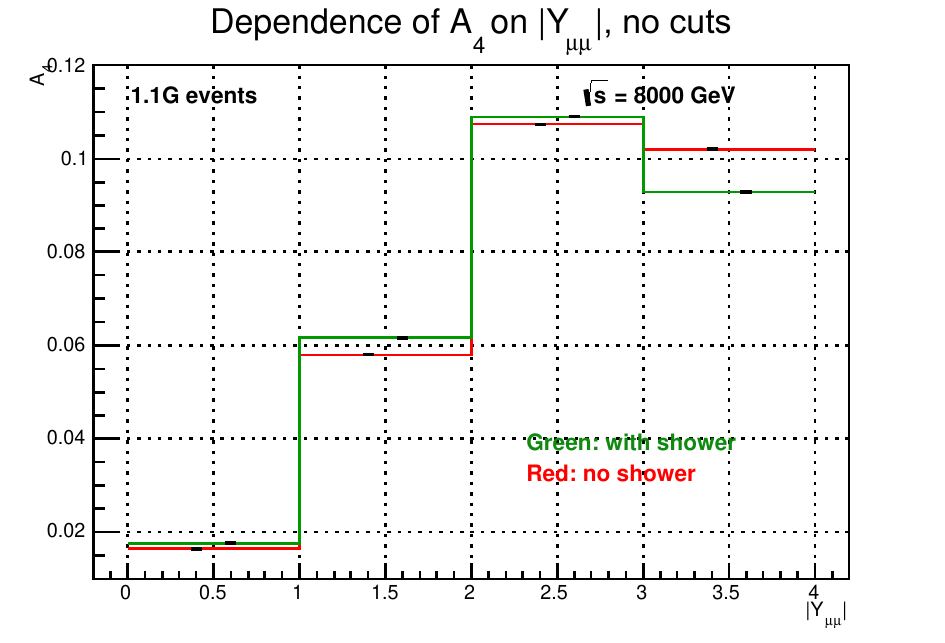}}
\put(3.2,.2){\includegraphics[width=3.2in,height=2.0in]{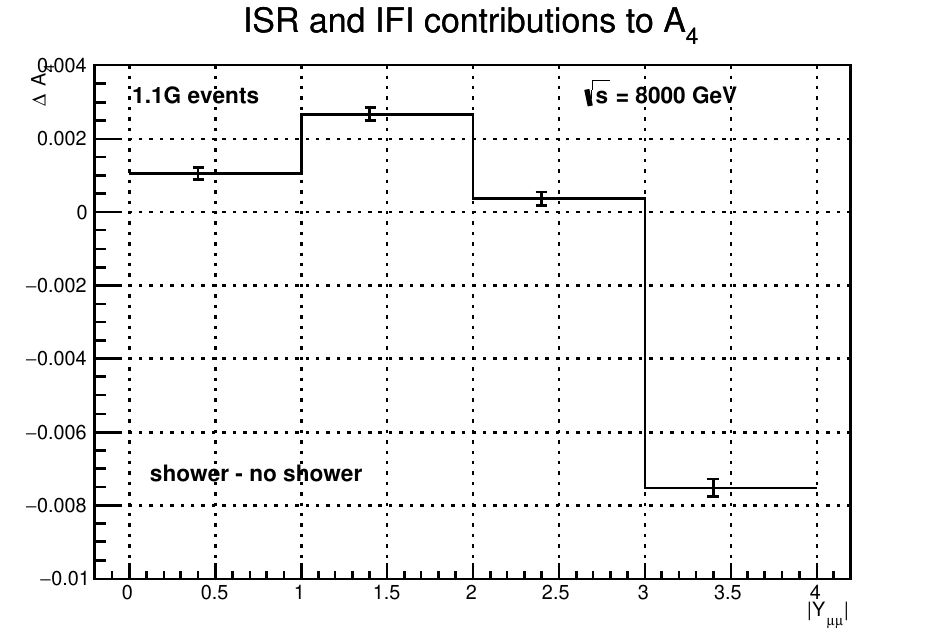}}
\end{picture}
\end{center}
\vspace{-10mm}
\caption{\baselineskip=11pt  Showered contributions to  $A_4$: the two top plots show the shower effects as a function of $M_{\ell\ell}$, the two bottom plots show the shower effects as a function of $Y_{\ell\ell}$. Showered {\KK}MC-hh ISR+ FSR+IFI results are shown in green (light shade), and unshowered {\KK}MC-hh ISR+ FSR+IFI results are shown in red (medium dark shade). In the two plots on the right,
the respective differences between the showered and unshowered distributions are shown.}
\label{fig12}
\end{figure}

In Fig.~\ref{fig12} we show the effects of the shower on $A_4$. 
The effect of the
shower on $A_4$   is small
for $M_{\ell\ell}\ge M_Z$. As a function the rapidity, the effect of the shower on $A_4$ is fairly small except for large values of $Y_{\ell\ell}$.
\par
\begin{figure}[h]
\begin{center}
\setlength{\unitlength}{1in}
\begin{picture}(6.5,3.0)(0,0)
\put(0,0.5){\includegraphics[width=3.2in,height=2.6in]{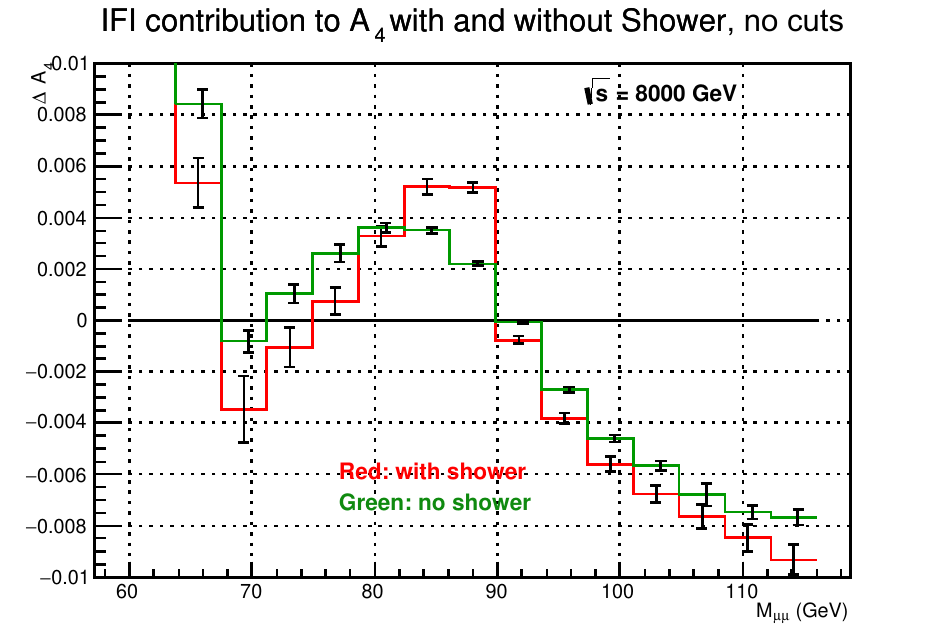}}
\put(3.0,0.5){\includegraphics[width=3.2in,height=2.6in]{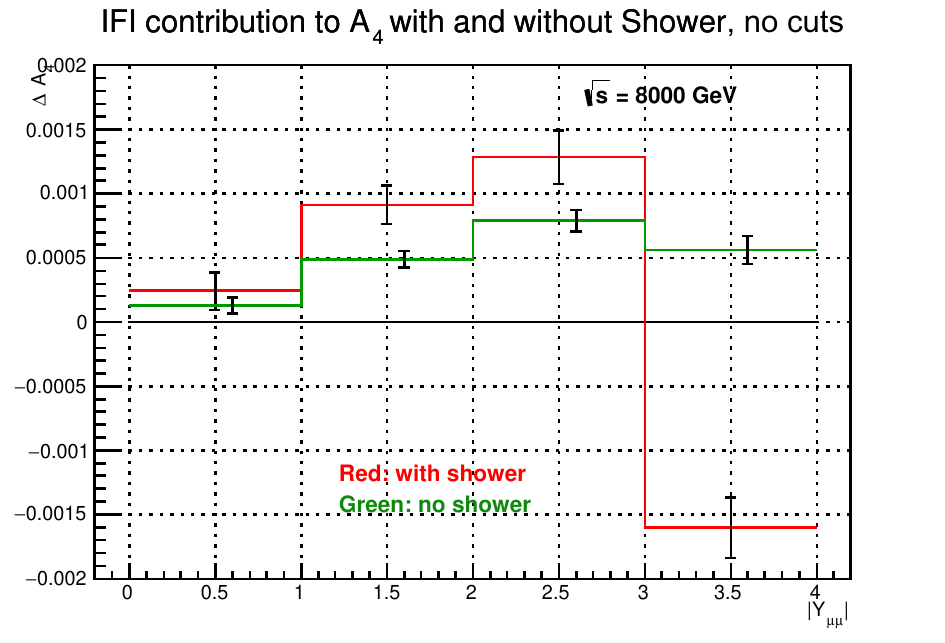}}
\end{picture}
\end{center}
\vspace{-10mm}
\caption{\baselineskip=11pt Shower effects on the IFI contribution to $A_4$:  the plots on the left illustrate the shower effects as a function of $M_{\ell\ell}$, the plots on the right illustrate the shower effects as a function of $Y_{\ell\ell}$. The results show the showered and unshowered differences between {\KK}MC-hh ISR+ FSR+IFI results and the corresponding {\KK}MC-hh ISR+ FSR results. The unshowered results are are in green (light shade), the showered results are in red (medium dark shade).}
\label{fig13}
\end{figure}

The shower effects on the IFI contribution to $A_4$ are studied in Fig.~\ref{fig13}.
There are modulations by the shower of both the distribution in $M_{\ell\ell}$ and the distribution in $Y_{\ell\ell}$. Precision studies should not ignore these effects as it is well-known.\par
\section{Summary}
Our results show that ISR typically enters the angular results ($A_{FB}$, $A_4$) at the level of several per mille. Both \KK{MC}-hh
and QED PDFs give a comparable ISR effect on angular results. The IFI effect is typically one-tenth the ISR effect or less, but this is sensitive to cuts.
The parton shower changes the detailed results, but not the general size of the ISR and IFI
corrections. A more complete treatment of the respective QCD corrections, accurate to NLO, in the presence of our exact ${\cal O}(\alpha^2 L)$ CEEX EW corrections will appear elsewhere~\cite{elswh}.\par
Studies~\cite{froid:2019} are underway to clarify the role of ISR and IFI in the precision determination of
$\sin^2\theta_W$ from LHC data. In these studies, approaches based on collinear QED PDF's\footnote{See 
Ref.~\cite{vicini-wack:2016} for a complete list of all the approaches which use collinear QED PDF's.} will be compared with the approach in KKMC-hh
to elucidate the relationship between the different approaches with the objective of defining the relevant theoretical precision tag.\par
We note the ISR in \KK{MC}-hh is sensitive to the value of light quark masses, as discussed in Ref.~\cite{kkmchh1}. The key point is that the results 
from Ref.~\cite{mstw-mass} show that the light quark masses must be the short distance type masses with $m_u \cong 6\; \text{MeV}, \; m_d \cong 10 \;\text{MeV}, m_s\cong 150\; \text{MeV}$
where the uncertainty may be estimated by taking the PDG~\cite{PDG:2016} values $m_u = 2.2\; \text{MeV}, m_d = 4.7\; \text{MeV}, m_s = 96\; \text{MeV}$. Since the quark masses enter via the big log $L_q= \ln(M_Z^2/m_q^2)$, we expect the fractional uncertainty in our results from such a change in our masses to be at the level of the weighted fractional change in $L_q$, which is $<\Delta L/L >\cong ((\frac{4}{9}\Delta L_u+\frac{1}{9}\Delta L_d)/((\frac{4}{9} L_u+\frac{1}{9}L_d)\cong 0.10,$ if use the fact that the densities of u and d quarks at the relevant momentum fractions are almost equal inside the proton. Here,
$\Delta L_q= \ln(M_Z^2/m_{q2}^2) - \ln(M_Z^2/m_{q1}^2)$ for the two masses $m_{qi},\; i=1,2,$ for quark q in an obvious notation.
Further studies on the role of light quark masses in EW higher order corrections in precision LHC/FCC physics are in progress and will appear elsewhere~\cite{elswh}.\par
Precision studies of angular observables in single $Z/\gamma^*$ production at the LHC must take the effects from EW ISR that we have discussed in this paper into account.\par

\vskip 2 mm
\centerline{\bf Acknowledgments}
\vskip 2 mm

This work was supported in part 
by the Programme of the French–Polish Cooperation between IN2P3 and COPIN 
within the Collaborations Nos. 10-138 and 11-142 and by a grant from the Citadel Foundation. The authors also thank the IFJ-PAN, Krakow, PL for computing support and Prof. G. Giudice for the support and kind hospitality of the CERN TH Department.

\bibliography{Tauola_interface_design}{}
\bibliographystyle{utphys_spires}












\end{document}